\documentclass[compsoc, conference, a4paper, 10pt, times]{IEEEtran}

\usepackage{subfigure}

\usepackage{amsmath,amsfonts,mathtools,mathrsfs}
\usepackage{amsthm}
\usepackage[sans]{dsfont}
\usepackage{stmaryrd}
\usepackage{pifont}
\usepackage{wasysym}

\usepackage{textcomp}

\usepackage[T1]{fontenc}

\usepackage{tabularx,adjustbox}
\usepackage{booktabs}
\usepackage{graphicx}  
\usepackage{algorithm,algpseudocode} 

\usepackage{multirow}

\usepackage{url}

\makeatletter
\newcommand{\vast}{\bBigg@{4}}
\newcommand{\Vast}{\bBigg@{5}}
\newcommand{\immense}{\bBigg@{6}}
\newcommand{\Immense}{\bBigg@{7}}
\newcommand{\IImmense}{\bBigg@{8}}
\makeatother

\usepackage{cite}
\usepackage{longfbox}

\usepackage{caption,graphicx,newfloat}

\usepackage{ragged2e}

\usepackage{parallel,enumitem}
\newcommand{\cmark}{\ding{51}}%
\newcommand{\xmark}{\ding{55}}%

\begin{document}

\title{Practically Efficient Secure Computation of Rank-based Statistics \\ Over Distributed Datasets}

\author{\IEEEauthorblockN{Nan Wang}
\IEEEauthorblockA{\textit{Australian National University} \\
vincent.wang@anu.edu.au}
\and
\IEEEauthorblockN{Sid Chi-Kin Chau}
\IEEEauthorblockA{\textit{Australian National University} \\
sid.chau@anu.edu.au}
}

\maketitle

\begin{abstract}
In this paper, we propose a practically efficient model for securely computing rank-based statistics, e.g., median, percentiles and quartiles, over distributed datasets in the malicious setting without leaking individual data privacy. Based on the binary search technique of Aggarwal et al. (EUROCRYPT \textquotesingle 04), we respectively present an interactive protocol and a non-interactive protocol, involving at most $\log ||R||$ rounds, where $||R||$ is the range size of the dataset elements. Besides, we introduce a series of optimisation techniques to reduce the round complexity. Our computing model is modular and can be instantiated with either homomorphic encryption or secret-sharing schemes. Compared to the state-of-the-art solutions, it provides stronger security and privacy while maintaining high efficiency and accuracy. Unlike differential-privacy-based solutions, it does not suffer a trade-off between accuracy and privacy. On the other hand, it only involves $O(N \log ||R||)$ time complexity, which is far more efficient than those bitwise-comparison-based solutions with $O(N^2\log ||R||)$ time complexity, where $N$ is the dataset size. Finally, we provide a UC-secure instantiation with the threshold Paillier cryptosystem and $\Sigma$-protocol zero-knowledge proofs of knowledge.

\end{abstract}

\section{Introduction} \label{sec:intro}

Over the past decades, vast amounts of data have been generated and collected from various distributed sources. Extracting the essential statistics can help us gain insight into the nature of these data. Highly susceptible to outliers, traditional statistics such as mean and variance do not provide helpful information when datasets are skewed, where outliers are values greatly different from most others. Rank-based statistics, e.g., median, percentiles, and quartiles, play a crucial role in analysing the patterns of distributed datasets. Less influenced by outliers, these rank-based statistics have extensive applications in real-world scenarios, greatly benefitting our daily lives. For example, the median price of the houses in a region is helpful in estimating the market value of a house. In transportation, the $85^{th}$ percentile speed value is usually used to establish the speed limit of a road \cite{85speed}. In children's growth charts, percentiles are illustrated to help parents learn their children's growth status. The interquartile range, namely, the difference between the $25^{th}$ and $75^{th}$ percentile, can tell us how widely distributed the values are in a dataset. Data privacy has drawn significant attention in recent years as people are growingly concerned about privacy leakage and data misuse. Hence, it is becoming increasingly difficult to efficiently, securely and accurately extract these vital statistics from distributed datasets. Current proposals either suffer from a trade-off between privacy and accuracy, or poor applicability and efficiency in real-world scenarios. In this paper, we explore a feasible solution to the secure computation of these rank-based statistics with high accuracy, efficiency, privacy and applicability by mitigating the limitations of current solutions.

\subsection{State-of-the-art Solutions}

We begin with brief descriptions and analyses of the state-of-the-art solutions before elaborating on our contributions. Please see Table \ref{tab:symbols} for global key symbols and notations throughout the paper. We firstly introduce two important security settings, the {\em semi-honest} setting and the {\em malicious} setting:
\begin{itemize}
\item The semi-honest setting assumes an ideal world where the parties honestly follow the protocols but are curious to infer privacy based on the communicated transcripts. The protocols tend to yield high computational efficiency but ensure weak security due to a disregard for malicious behaviours, which are less practical in real-world scenarios. 

\item The malicious setting simulates real-world circumstances where the parties may arbitrarily deviate from the prescribed protocols. The protocols must defend against all kinds of undesirable behaviours of malicious parties.

\end{itemize}

Current solutions are essentially based on three core techniques, namely, multi-party computation (MPC), homomorphic encryption (HE) and differential privacy (DP). They handle the following two scenarios in different security settings:
\begin{itemize}
\item{\em Union of multiple users' individual data} (Scenario-I): There is a group of users $({\mathcal{U}_i})_{i=1}^N$, where each ${\mathcal{U}_i}$ holds an input $x_i$.

\item{\em Union of multiple datasets} (Scenario-II): There is a group of users $({\mathcal{U}_i})_{i=1}^N$, where ${\mathcal{U}_i}$ holds a dataset $D_i$. 
\end{itemize}

\begin{table}[!h]
\centering
\caption{Table of key symbols and notations.}
\resizebox{0.48\textwidth}{!}{%
\begin{tabular}{@{}cl@{}}
\toprule
   $({\mathcal{U}_i})_{i=1}^N$ & A group of $N$ users \\
   $({\mathcal{W}_j})_{j=1}^J$ & A group of $J$ worker servers \\
   $x_i$ & The input of the $i$-th user \\
   $\langle x_i \rangle$ & The encoded input of the $i$-th user \\
   $\underline{B}, \overline{B}$ & The lower and upper bounds of the inputs \\
   $R=\overline{B}-\underline{B}$ & The public range of the inputs \\
   $||\cdot||$ & The cardinality of a value space \\
   $|\cdot|$ & The absolute value \\
   $\alpha, \beta$ & The lower and upper bounds of a range \\
   $m, \mu, \sigma$ & Median, mean, standard deviation \\
   $D$ & A dataset \\
   $c$ & The number of colluding worker servers \\
   $\eta$ & The scaling factor \\
   {\rm Scenario-I} & Union of multiple users' individual data \\ 
   {\rm Scenario-II} & Union of multiple datasets \\
   {\rm ZKPoK} & A zero-knowledge proof of knowledge \\
   ${\mathcal L}$ & The NP language of a ZKPoK \\
   {\rm PPT} & Probabilistic polynomial time  \\ \bottomrule
\end{tabular}%
}
\label{tab:symbols}
\end{table}

\subsubsection{MPC-Based Computation}

Aggarwal et al. (EUROCRYPT \textquotesingle 04) \cite{mpcmedian} proposed a binary-search-based MPC protocol to securely compute the $k^{th}$ element for Scenario-II in the malicious setting. This approach can yield the actual median with high accuracy. Intuitively, the users initialise a search range $[\alpha=\underline{B}, \beta=\overline{B}]$ and recursively lessen the range based on guessed values to finalise the target value within at most $\lceil \log (||R||) \rceil$ rounds, where $R=[\underline{B}, \overline{B}]$ is the public range of the elements in the datasets. We describe each round of the protocol for the median computation as follows: 
\begin{enumerate}
\item a guessed median value $m=\lceil \frac{\alpha+\beta}{2} \rceil$ is suggested.

\item ${\mathcal{U}_i}$ broadcasts a value $l_i$, the number of values smaller than $m$, and a value $g_i$, the number of values larger than $m$. The users compare the sum $s=\sum l_i$ with the middle value $\frac{\sum ||D_i||}{2}$.

\item the following operations will be performed:
\begin{align*}
\begin{split}
	\left\{ 
	\begin{array}{ll}
		{\rm terminate}, & \mbox{if \ } s=\frac{\sum ||D_i||}{2} \\
		\alpha \leftarrow m,(g)^i=||D_i||-l_i & \mbox{if \ } s<\frac{\sum ||D_i||}{2} \\
		\beta \leftarrow m,(l)^i=||D_i||-g_i & \mbox{if \ } s>\frac{\sum ||D_i||}{2}
	\end{array}\right.
\end{split}
\end{align*}
where $l^{(i)}$ and $g^{(i)}$ indicate the number of inputs strictly greater and smaller than the current range $[\alpha, \beta]$. 
\end{enumerate}
If $s=\frac{\sum ||D_i||}{2}$, the protocol terminates with $m$ being the median because the number of elements smaller than $m$ is equal to $\frac{\sum ||D_i||}{2}$. If $s < \frac{\sum ||D_i||}{2}$, it means the number of elements smaller than $m$ is less than $\frac{\sum ||D_i||}{2}$, indicating that the actual median lies in $[m, \beta]$. Otherwise, if $s > \frac{\sum ||D_i||}{2}$, it means the actual median lies in $[\alpha, m]$. The users restart from step 1 to search the median in smaller ranges. To prevent any user from providing inconsistent data throughout the protocol, the users are allowed to perform consistency checks by comparing $l_i+g_i \leq ||D_i||, l_i \geq l^{(i)}, g_i \geq g^{(i)}$. 

The solution is highly efficient but has four weaknesses. First, the local distribution privacy $l_i$ and $g_i$ of the dataset $D_i$ must be leaked to achieve the global median search. Second, the consistency checks are less rigorous and still leave room for cheating, which may result in less accurate results. Third, the protocol is not general for all types of distributed datasets. For Scenario-I, individual data privacy would be disclosed to obtain $l_i$ and $g_i$ to adjust the search range. Fourth, the protocol requires all the users to remain online and deeply involved, which is less friendly to those with limited computational and communication resources.

\subsubsection{HE-Based Computation}

The two recent studies \cite{blockchainmedian, starmedian} based on homomorphic encryptions (HE) were conducted to achieve the secure computation of rank-based statistics for Scenario-I with constant-round protocols. Both studies adopted the secure bitwise comparison method based on the DGK technique \cite{DGK}. The method requires each user to submit $\log ||R||$ encrypted bit values of the input for $N-1$ integer comparisons with other users' inputs, where $N$ is the number of users and $R$ is the range of the inputs. In total, it demands $O(N^2 \log ||R||)$ computationally expensive decryptions for $O(N^2)$ comparisons. Although it is argued that the computations can be performed in parallel to decrease the running time, the demanded computational resources remain unchanged. Moreover, their solutions do not scale well with a large number of users. For instance, handling 1,000 users alone would require a staggering one million comparisons.

The major difference is that the former work (ASIACCS \textquotesingle 20) \cite{blockchainmedian} uses a non-trivial $O(N^2 \log ||R||)$ number of zero-knowledge proofs to prove the correctness of re-encryptions, comparisons, integer shuffling, etc., in the malicious setting. However, the privacy of whether a user's input is smaller than others' inputs is not well preserved during the comparison. The latter (FC \textquotesingle 20) \cite{starmedian} employs a client-server communication model to outsource computations to a central server. Their work presented three approaches based on garbled circuits, additively homomorphic encryption (AHE) and somewhat homomorphic encryption (SHE) in the semi-honest setting. Notably, the SHE-based approach provides non-interactivity support, where the server performs computations without users' involvement. Unfortunately, their SHE-based method is still not computationally efficient even without using zero-knowledge proofs. In their experiment, it took almost an hour to process 25 users' data.

\subsubsection{DP-Based Computation}

Two differential-privacy-based solutions \cite{differentialmedian1, differentialmedian2} were presented to compute the median over distributed datasets. Both solutions combine the exponential mechanism technique and secret-sharing schemes to achieve high computational efficiency in the semi-honest setting. However, both solutions have three major limitations. First, the solutions are subject to the inherent limitation of the DP technique that a trade-off between privacy and accuracy has to be made by adjusting a privacy hyperparameter $\epsilon \in [0, 1]$. The larger $\epsilon$, the more accurate the result, the less private the data and vice versa. For instance,  the authors \cite{differentialmedian1} adopted a technique that divides the datasets into $S$ subranges. However, there is at least $1-\zeta$ chance that their algorithm outputs an element that is at most $\frac{\ln (S/\zeta)}{\epsilon}$ positions away from the actual median. Even if $\epsilon$ is set to 1 in disregard of the privacy, the most accurate value their algorithm can yield is still at most 7 positions away from the actual median when $S=10$ and $\zeta=0.01$, where $S=10$ is the optimal value in their experiments. Thus, DP-based solutions are not desirable for scenarios with both high privacy and accuracy requirements. Second, their solutions are only proven secure in the semi-honest setting without considering dishonest behaviours\footnote{The protocols and experiments in both papers \cite{differentialmedian1, differentialmedian2} were based on the semi-honest setting. A brief extension to the malicious setting was provided in \cite{differentialmedian1}, but no rigorous security proofs were given.}. Third, the same as \cite{mpcmedian}, the solutions require users' interaction to jointly compute the statistics.

More concretely, the former solution (USENIX-Security \textquotesingle 20) \cite{differentialmedian1} adopted a client-server communication model to handle Scenario-I, where the users are allowed to outsource computations to several semi-honest, non-colluding servers. Nevertheless, on the one hand, it incurs high round complexity for involving a non-trivial number of multi-round sub-protocols to achieve basic secure computations. On the other hand, it does not withstand collusion attacks mounted by the servers. The latter (NDSS \textquotesingle 20) \cite{differentialmedian2} is designed for Scenario-II, which only works with two distributed datasets by leveraging garbled circuits to achieve secure comparisons. It remains unknown whether it applies to multiple distributed datasets.

\subsection{Contributions}

In this paper, we propose a practically efficient model for securely computing rank-based statistics, e.g., median, percentiles and quartiles, over distributed datasets in the malicious setting without leaking individual data privacy. Inspired by the binary search idea \cite{mpcmedian}, we respectively present an interactive protocol $\Pi_{\sf irank}$ and a non-interactive protocol $\Pi_{\sf nirank}$ in a client-server communication model, where a group of users outsource computations to a set of servers by leveraging their powerful computational and communication resources. $\Pi_{\sf irank}$ involves users' intervention to provide intermediate results. It demands less computational workload on the server side but is vulnerable to the early-quit attack, where the users may intentionally or unintentionally quit amid the protocol. The early-quit attack would undermine the computed statistics' accuracy, damaging to all kinds of interactive protocols. Thus, we propose a distributed masking scheme that allows to remove users' intervention without compromising individual privacy. The protocol $\Pi_{\sf nirank}$ can withstand the early-quit attack but requires more server-side computational and communication resources. 

\subsubsection{Intuition}

Our computing model requires users to encode their inputs using homomorphic encryption or secret-sharing schemes so that all the computations can be performed on encoded data to preserve privacy. The current high-accuracy solutions either explicitly reveal individual privacy,  e.g., the number of inputs smaller/larger than guessed values, or demand computationally expensive comparisons of all the users' inputs. Unlike them, our solution benefits from a new secure {\em sign extraction} technique that allows to extract the sign of $x_i-m$, either 1 or -1, without leaking the individual privacy of users' data. Then all the signs will be securely added up before being compared with 0 to adjust the search range accordingly. The same as the MPC-based solution \cite{mpcmedian}, the maximum round complexity is logarithmic in the range size of the users' inputs. We also introduce a series of optimisation techniques to reduce the round complexity.

\subsubsection{Comparison}

Compared with the state-of-the-art solutions, our computing model:
\begin{itemize}
\item is more secure and applicable in real-world scenarios. It only reveals the global knowledge, e.g., rank-based statistics and global distribution of distributed datasets, without leaking individual privacy, where privacy refers to a user's data or the local distribution of a user's dataset. Besides, our computing model can defend against collusion attacks.

\item achieves superior computational efficiency than the bitwise-comparison-based solutions \cite{blockchainmedian, starmedian} with $O(N^2\log ||R||)$ time complexity. Ours only requires $O(N\log ||R||)$ time complexity. More concretely, the solution  \cite{blockchainmedian} requires $O(N^2 \log ||R||)$ zero-knowledge proofs to defend against malicious behaviours, whereas ours only demands $O(N\log ||R||)$ zero-knowledge proofs.

\item supports non-interactivity, which relieves users of expensive computations by outsourcing them to powerful servers
without having to keep users online. The non-interactive protocol is highly friendly to those users with limited resources. Compared to the solely non-interactive solution \cite{starmedian} with $O(N^2\log ||R||)$ time complexity, our solution presents a considerably greater advantage in computational efficiency.

\item yields high accuracy and privacy simultaneously without suffering from a trade-off.

\item applies to two general distributed scenarios and especially works with multiple distributed datasets.

\item is modular and can be instantiated with either homomorphic encryption or secret-sharing schemes. 

\end{itemize}
In summary, to the best of our knowledge, our solution achieves the best overall performance in the malicious setting at the time of writing with respect to efficiency, privacy, security\footnote{Security refers to the malicious setting and collusion.}, accuracy, generality and practicality\footnote{Our solution is less computationally efficient than those not requiring zero-knowledge proofs or in the semi-honest setting.}. Finally, we provide a UC-secure (Universal Composability)\footnote{Universal composability framework is a stronger malicious model than the stand-alone model. Please refer to Section \ref{sec:ucmodel} for more details.} instantiation of our computing model using the threshold Paillier cryptosystem and a tuple of $\Sigma$-protocol zero-knowledge proofs of knowledge. The instantiation is proven secure against static, active (malicious) adversaries, where {\em static} means that the adversaries can perform corruptions at the outset of the protocols. 

\subsection{Outline of the paper}

The paper is organised as follows. First, we introduce the preliminaries in Section \ref{sec:preliminaries}. We elaborate on the two protocols of securely computing rank-based statistics for Scenario-I and some optimisations in Section \ref{sec:protocol1} and \ref{sec:protocol2}. Then we briefly depict the secure computation of Scenario-II in Section \ref{sec:protocol2}. We provide an instantiation of our protocols with the threshold Paillier cryptosystem and describe the security proofs in Section \ref{sec:instantiation}. A performance evaluation is given in Section \ref{sec:experiment}.

\section{Preliminaries} \label{sec:preliminaries}

\subsection{Homomorphic Enc-Dec Scheme}

We define an abstract homomorphic Enc-Dec scheme consisting of two PPT algorithms, namely, encode and decode:
\begin{itemize}
\item The encode algorithm $\langle \theta \rangle \leftarrow {\rm Enc}(\theta)$ takes an integer plaintext $\theta \in \Theta$ and outputs a ciphertext $\langle \theta \rangle$, where $\Theta$ is the plaintext space. 

\item The decode algorithm $\theta \leftarrow {\rm Dec}(\langle \theta \rangle)$ takes a ciphertext $\langle \theta \rangle$ and outputs a plaintext $\theta \in \Theta$.

\end{itemize}
We require the scheme to be {\em hiding} if $\langle \theta \rangle$ does not reveal $\theta$, and {\em binding} if $\langle \theta \rangle$ can only be decoded to $\theta$. The Enc-Dec scheme should be homomorphic and satisfies the following properties:
\begin{align*}
\begin{split}
\langle \theta \rangle \oplus \langle \delta \rangle = \langle \theta+\delta \rangle
\end{split} &
\begin{split}
\langle \theta \rangle \ominus \langle \delta \rangle = \langle \theta-\delta \rangle
\end{split} \\
\begin{split}
\langle \theta \rangle \otimes \delta = \langle \theta \cdot \delta \rangle
\end{split} &
\begin{split}
\langle \theta \rangle \otimes \langle \delta \rangle = \langle \theta \cdot \delta \rangle
\end{split} 
\end{align*}
where $\theta$ and $\delta$ are two plaintext values. $\oplus$, $\ominus$, $\otimes$ indicate three operations on encoded data.

The Enc-Dec scheme only supports integer values. Thus, we use the techniques in \cite{wang2012harnessing,miao2015cloud} to expand its usage with fractional numbers. The basic idea is to choose a proper scaling factor $\eta$ to enlarge fractional numbers to integers. After decoding, the actual results can be obtained by scaling down by $\eta$. This naturally retains all the effective digits of fractional numbers without compromising accuracy. Moreover, all the relevant terms in the computations must also be scaled up by the same scaling factor to yield correct results.

\subsection{Zero-Knowledge Proofs of Knowledge}

A zero-knowledge proof of knowledge (ZKPoK) allows a prover ${\mathcal{P}}$ to convince a verifier ${\mathcal{V}}$ of knowing a secret without leaking any information. More formally, given an NP-language ${\mathcal{L}}$, a prover can convince a verifier of knowing a witness for a statement ${\rm stmt} \in {\mathcal{L}}$. In our paper, we leverage $\Sigma$-protocol ZKPoKs to compel parties to follow the prescribed protocols. A $\Sigma$-protocol goes as follows:
\begin{enumerate}
\item ${\mathcal{P}}$ sends an initial message to ${\mathcal{V}}$.

\item ${\mathcal{V}}$ replies with a uniformly random challenge. 

\item ${\mathcal{P}}$ responds to the challenge. 
\end{enumerate}
Finally, ${\mathcal{V}}$ decides whether to accept or reject the proof.

\noindent
Next, we briefly describe five useful ZKPoKs:
\begin{itemize}

\item{\em ZKP of Multiplication:} Given $\langle \theta \rangle $, $\langle \delta \rangle$ and $\langle \rho \rangle$, ${\mathcal{P}}$ can convince ${\mathcal{V}}$ of $\rho=\theta \cdot \delta$. The proof is denoted by zkpMTP$\{\langle \theta \rangle, \langle \delta \rangle, \langle \rho \rangle\}$. This proof can also prove that the multiplication is correct without knowing $\theta$.

\item{\em ZKP of Membership:} Given a public set $\Theta = \{\theta_1, ..., \theta_T\}$ and $\langle \theta \rangle$, ${\mathcal{P}}$ can convince ${\mathcal{V}}$ of $\theta \in \Theta$. The proof is denoted by zkpMBS$\{\langle \theta \rangle\}$. 

\item{\em ZKP of Range:} Given $\langle \theta \rangle$, ${\mathcal{P}}$ can convince ${\mathcal{V}}$ of $\theta \in [\alpha, \beta]$. The proof is denoted by zkpRG$\{\langle \theta \rangle, [\alpha, \beta]\}$.

\item{\em ZKP of Non-Zero:} Given $\langle \theta \rangle $, ${\mathcal{P}}$ can convince ${\mathcal{V}}$ of $\theta \neq 0$. The proof is denoted by zkpNZ$\{\langle \theta \rangle, \theta \neq 0\}$. 

\item{\em ZKP of Decode:} Given $\langle \theta \rangle $, ${\mathcal{P}}$ can convince ${\mathcal{V}}$ that $\theta$ is the correct decoded value. The proof is denoted by zkpDEC$\{\langle \theta \rangle\}$. 

\end{itemize}
Moreover, $\Sigma$-protocol ZKPoKs can be made non-interactive via Fiat-Shamir heuristic \cite{FiatShamir} with the use of a collision-resistant hash function ${\mathcal H}(\cdot)$ modelled as a random oracle \cite{randomoracle}. The non-interactive ZKPoKs (NIZKPoKs) fulfils {\em completeness}, {\em soundness} and {\em zero-knowledge} in the random oracle model. Please refer to Appendix \ref{sec:append1} for concrete protocols.

\section{Security Assumption}

In this paper, we essentially focus on the secure computation of Scenario-I. 

\subsection{Communication Model} 

\begin{figure}[!h]
\centering
\includegraphics[width=0.34\textwidth]{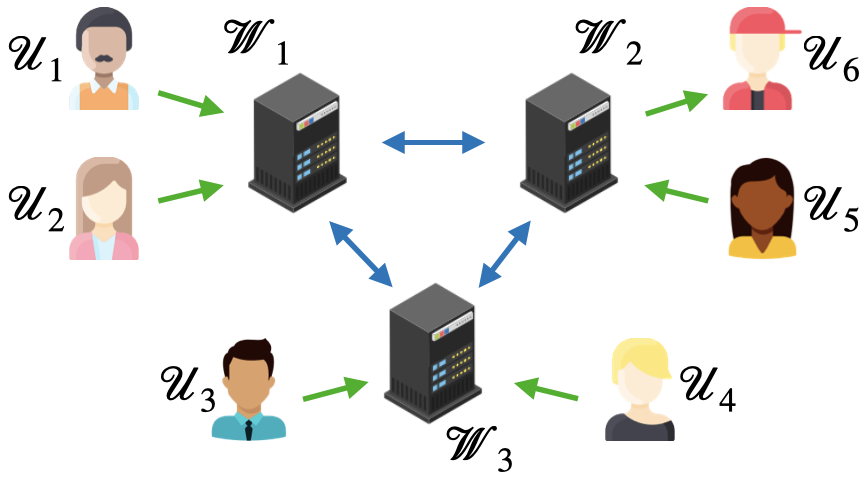}
\caption{An example of the client-server communication model with 6 users and 3 worker servers.}
\label{fig:diagram}
\end{figure}

We adopt the same client-server communication model as \cite{differentialmedian1} shown in Figure \ref{fig:diagram}, where a group of $N$ users $({\mathcal{U}_i})_{i=1}^N$ and a set of $J$ worker servers\footnote{The servers can be different cloud computing platforms such as Amazon Web Services, Google Cloud, Microsoft Azure, etc.} $({\mathcal{W}}_j)_{j=1}^J$ jointly perform secure computations. The client-server model is more promising than the traditional MPC model, where the latter allows the users to communicate with each other via peer-to-peer channels. Generally, users' locations tend to be scattered across different regions and geographically distant from each other. The peer-to-peer channels greatly increase the communication overhead, causing high network latency. Instead, the client-server model allows the users to communicate with their geographically nearest servers, which takes advantage of the high-speed and high-bandwidth network to achieve faster and more stable communication.

\subsection{Threat Model} 

Each user ${\mathcal{U}_i}$ holds an integer input $x_i$ in a public range $R=[\underline{B}, \overline{B}]$, where $\underline{B}$ and $\overline{B}$ indicate the lower and upper bounds of the inputs. We assume an honest majority of users, where less than $\dfrac{N}{2}$ users are malicious and deviate arbitrarily from the protocols. Malicious users may:
\begin{itemize}
\item submit outliers 

\item early quit the protocol

\item provide inconsistent data

\item collude together

\end{itemize}
The current solutions do not commonly consider outliers and early-quit attacks. However, these misconducts are not uncommon in real-world scenarios. We assume the existence of a small portion of outliers. On the one hand, rank-based statistics are inherently insensitive to outliers. On the other hand, we use zero-knowledge range proofs to confine users' inputs in the allowed range $R$, such that the outliers would not heavily bias the computed statistics. Furthermore, we consider the early-quit attack that is fairly destructive to interactive protocols rather than non-interactive ones. There are several workarounds to mitigate the damage to interactive protocols. For instance, penalties can be used to prevent users from quitting early. Each user must make a deposit beforehand, which would be confiscated if the user early quits the protocol. 

We assume $J$ worker servers, who jointly perform aggregation of the users' encoded inputs and decode the aggregated results. In the following, we will call them {\em workers} for brevity. Besides, we assume $c$ out of $J$ workers may act maliciously, where $c \in \{0, 1, ..., J-1\}$. They may:
\begin{itemize}

\item refuse to follow the protocols

\item collude together to pry into users' privacy

\end{itemize}

We assume a worst-case monolithic adversary that can control all the malicious users and servers. Our computing model uses zero-knowledge proofs to achieve the identifiable abort, where at least one honest worker aborts the protocols in case of detecting any malicious acts so as to prevent any privacy leakage. Likewise, deposit penalties can also be used to prevent any malicious workers from sabotaging the protocols. 

\section{Interactive Protocol $\Pi_{\sf irank}$} \label{sec:protocol1}

The interactive protocol entails users' involvement, where the users must remain online throughout the protocol, performing the prescribed computations. We begin with the protocol to securely compute the median before presenting the generalised version of computing the $k^{th}$ element. We define a sign extraction function $\phi(\theta)$ that extracts the sign of a non-zero input $\theta$ below:
\begin{align*}
\begin{split}
	\phi(\theta) =&  \left\{ 
	\begin{array}{rl}
		1, & \mbox{if \ } \theta > 0 \\
		-1, & \mbox{if \ } \theta < 0
	\end{array}\right.
\end{split}
\end{align*}

\subsection{Median Computation Protocol} 

We stress that our protocol only reveals the global knowledge of distributed datasets, e.g., median and global distribution, without leaking individual data privacy. Before the protocol, the users make a consensus on a binary search algorithm ${\mathcal{B}}$ to dynamically update the search range throughout the protocol. Each user ${\mathcal{U}_i}$ submits an encoded integer $\langle x_i \rangle$ as the initial input along with a zkpRG$\{\langle x_i \rangle, R=[\underline{B}, \overline{B}]\}$ to the workers ${\mathcal{W}}$. The protocol involves, at most, a logarithmic number of rounds, and the $l$-th round goes as follows, where $l \in \{1, 2, ...\}$:

\begin{enumerate}
\item ${\mathcal{B}}$ suggests a fractional median $m_l=\lfloor \dfrac{\alpha+\beta}{2} \rfloor + \dfrac{1}{2}$ for a search range $[\alpha,\beta]$ and encodes it as $\langle \eta \cdot m_l \rangle$, where $\alpha$ and $\beta$ are initialised as $\underline{B}$ and $\overline{B}$, respectively. Then each ${\mathcal{U}_i}$ computes an encoded value $\langle q_i \rangle$:
$$\langle q_i \rangle = \langle x_i \rangle \otimes {\eta}  \ominus \langle \eta \cdot m_l \rangle  = \langle \eta \cdot (x_i-m_l) \rangle$$
where $\eta$ is a large scaling factor.

${\mathcal{U}_i}$ submits the encoded value $\langle \phi(q_i) \rangle$ to ${\mathcal{W}}$ with a ZKPoK[${\mathcal{L}}_1$] for an NP language ${\mathcal{L}}_1$ proving the consistency between $\langle q_i \rangle$ and $\langle \phi(q_i)\rangle$, where $\phi(q_i) \in \{-1, 1\}$.

{\bf Remarks:} We add an extra $\dfrac{1}{2}$ to make $m_l$ fractional so that $q_i$ would never be zero to leak privacy. Besides, the correctness of $\langle q_i \rangle$ can be directly checked by following the formula without using zero-knowledge proofs.

\item ${\mathcal{W}}$ check the proofs and aggregate all the $\langle \phi(q_i)\rangle$ as $\sum_{i=1}^{N,\oplus}\langle \phi(q_i) \rangle$. Then ${\mathcal{W}}$ perform the decode and produce a zkpDEC.
$$z={\rm Dec}\big(\sum_{i=1}^{N,\oplus}\langle \phi(q_i) \rangle\big)=\sum_{i=1}^N \phi(q_i)$$ 
If $z=0$, the protocol terminates, and $\lceil m_l \rfloor$ is the target median. Otherwise, ${\mathcal{B}}$ updates the search range based on the rules on lines 5-9 in Algorithm \ref{alg:median} and proceeds to step (1). If $z$ is positive, the number of inputs larger than $m_l$ is greater than those smaller than $m_l$, and the actual median lies in the range $\big[\lfloor m_l \rfloor, \beta\big]$. If $z$ is negative, the number of inputs larger than $m_l$ is less than those smaller than $m_l$, and the actual median lies in the range $\big[\alpha, \lfloor m_l \rfloor\big]$. When $\beta - \alpha \leq 2$, the protocol terminates finally at the $(\lceil \log ||R|| \rceil-1)$-th round. 
\end{enumerate}

\begin{center}
\begin{minipage}{0.4\textwidth}
\begin{algorithm}[H]
\caption{\rm Secure Median Computation}\label{alg:median}
\begin{algorithmic}[1]
		\Loop
			\State $\textbf{execute}$ step (1), (2)
			\If{$z = 0$}
				\State \Return $\lceil m_l \rfloor$  
			\ElsIf{$z > 0$}
				\State $\alpha \gets \lfloor m_l \rfloor$
			\Else \Comment{{\em When $z<0$}}
				\State $\beta \gets \lfloor m_l \rfloor$
			\EndIf
			\If{$\beta - \alpha \leq 2$} 
				\State \Return $\lceil \dfrac{\alpha+\beta}{2} \rfloor$ \EndIf
		\EndLoop
	\end{algorithmic}
\end{algorithm}
\end{minipage}
\end{center}
{\bf Remarks:} The absolute value of $z$ tends to be smaller and smaller before gradually approaching 0. However, $z$ may not always end up being 0 when there are a number of duplicates around the actual median. It will oscillate till $\beta - \alpha \leq 2$ rather than gradually converge. Moreover, $z$ represents the global distribution knowledge, which does not leak individual data privacy. Besides, the rounding operations in the algorithm have a negligible impact on the final results. Users may choose whichever is suitable based on the actual scenarios. 

\subsection{ZKPoK[${\mathcal{L}}_1$]}

We use ZKPoKs to force the users to submit consistent values throughout the protocol. We formalise the language ${\mathcal{L}}_1$ of the ZKPoKs in step (1) of the protocol. Note that any zero-knowledge proofs satisfying the relations in ${\mathcal{L}}_1$ can be applied to our protocols. We will provide an instantiation of the $\Sigma$-protocol zero-knowledge proofs in Section \ref{sec:zkpinstantiation}. Let ${\rm stmt}$ and ${\rm wit}$ be the statement and witness.
$${\mathcal{L}}_1:=
\begin{cases}
{\rm stmt}=\big(\langle q_i \rangle, \langle \phi(q_i) \rangle, \langle \phi(q_i) \cdot q_i\rangle\big):\\
\exists {\rm wit}=\big(q_i, \phi(q_i)\big)~s.t.: \\
(1)~\phi(q_i) \in \{-1, 1\}\\
(2)~\phi(q_i) \cdot q_i > 0\\
\end{cases}
$$
The two witnesses $q_i$ and $\phi(q_i)$ must be subject to the above two constraints. Constraint (1) is straightforward. Constraint (2) requires $\phi(q_i)$ and $q_i$ to have the same sign based on the principle that the product of any two non-zero values with the same sign is always positive. 

\subsection{Optimisations}

Our computing model requires, at most, a logarithmic number of rounds in the range size of inputs. However, for large inputs, the round complexity would be non-trivial, leading to high network latency and running time. The user experience of the interactive protocol would be heavily undermined as the involved users have to stay online. Thus, we present three optimisation techniques and one optimisation technique to reduce the round complexity and the total running time, respectively.

\subsubsection{Optimisation-I}

To reduce the round complexity, the users can add an early terminating rule in ${\mathcal{B}}$ that the protocol terminates when $z \in [-\Delta, \Delta]$ rather than $z=0$, where $\Delta$ is a small fault-tolerant error determined according to the actual scenarios. Note that $\Delta$ should be discreetly chosen since large values may undermine the accuracy of the computed statistics.

\subsubsection{Optimisation-II}

In step (1), ${\mathcal{U}}_i$ provides an encoded sign value $\langle \phi(q_i) \rangle=\langle \phi\big(\eta \cdot (x_i -m_l)\big) \rangle$ for the $l$-th round. We introduce a technique to further reduce the round complexity at the expense of tripling communication costs. The users can pre-compute two possible encoded values of the subsequent round due to the predictability of the suggested values. For example, for a search range $[0, 100]$, the guessed median is initialised as $50.5$ in the first round and would be either 25.5 or 75.5 in the second round. Thus, the users can provide three encoded values $\langle \phi\big(\eta \cdot (x_i -50.5)\big) \rangle$, $\langle \phi\big(\eta \cdot (x_i -25.5)\big) \rangle$ and $\langle \phi\big(\eta \cdot (x_i -75.5)\big) \rangle$ in the first round. This optimisation technique helps halve the round complexity to $\dfrac{1}{2} (\lceil \log||R|| \rceil-1)$. Moreover, if the users have adequate computational and communication resources, they can pre-compute more possible encoded values to further reduce the round complexity.

\subsubsection{Optimisation-III}

In 1979, Book and Sher \cite{mediandistance} firstly proved the inequality that the distance between the mean $\mu$ and the median $m$ is not greater than one standard deviation $\sigma$ for a data distribution with the finite $\sigma$:
$$|\mu-m| \leq \sigma \Longleftrightarrow m \in [\mu-\sigma, \mu+\sigma]$$
Thus, given $\mu$ and $\sigma$, the guessed median can be initialised to $\mu$ in the first round, and the initial search range would be reduced to $[\mu-\sigma, \mu+\sigma]$. The maximum round complexity would be reduced to $\lceil \log \sigma \rceil$.

We briefly describe a protocol for securely computing $\mu$ and $\sigma$. Apart from the input $\langle x_i \rangle$, ${\mathcal{U}}_i$ must also provide an extra input $\langle x_i^2 \rangle=\langle x_i \rangle \otimes \langle x_i \rangle$. Then ${\mathcal{W}}$ can aggregate all the users' inputs by computing $S_x=\sum_{i=1}^{N,\oplus} \langle x_i \rangle$ and $S_{x^2}=\sum_{i=1}^{N,\oplus} \langle x_i^2 \rangle$. Next, ${\mathcal{W}}$ computes the $1^{st}$- and $2^{nd}$-order raw moments $\mu_x=\dfrac{S_x}{N}$ and $\mu_{x^2}=\dfrac{S_{x^2}}{N}$. Finally, the users can obtain $\sigma_x$ based on the formula $\sigma_x^2=\mu_{x^2}-\mu_x^2$. Note that one can also use this protocol to compute the $3^{rd}$- and $4^{th}$-order raw moments $\mu_{x^3}=\dfrac{S_{x^3}}{N}$ and $\mu_{x^4}=\dfrac{S_{x^4}}{N}$ in order to obtain the skewness $\gamma_x$ and kurtosis $\kappa_x$ of distributed datasets based on the formulae \cite{momentformulae} as below, where skewness reflects the asymmetry feature of a dataset and kurtosis is a measure of the "tailedness", which is a good indicator of the presence of outliers. Note that the sum values and raw moments are all global knowledge of distributed datasets.
\begin{align*}
\begin{split}
\gamma_x = \dfrac{\sum (x - \mu_x)^3}{N \sigma_x^3} = \dfrac{\mu_{x^3}-3  \mu_x \mu_{x^2} + 2  \mu^3_x}{\sigma^3_x}
\end{split} \\
\begin{split}
\kappa_x = \dfrac{\sum (x - \mu_x)^4}{N  \sigma_x^4} = \dfrac{\mu_{x^4}-4 \mu_x  \mu_{x^3}+6  \mu^2_x  \mu_{x^2}-3 \mu^4_x}{\sigma^4_x}
\end{split}
\end{align*}

\subsubsection{Optimisation-IV}

Our protocol leverages the ZKPoKs to defend against malicious users. However, verifying these proofs incurs non-trivial running time during the execution of the protocol, especially for those scenarios with large datasets but inadequate computational resources. Thus, to mitigate this issue, the workers can split the verification workload during the execution and perform the full verification after the execution. Concretely, each ${\mathcal{W}}_j$ is required to make a certain amount of deposit to a smart contract beforehand, where smart contracts are publicly verifiable computer programs that can automatically execute the agreements on blockchain platforms without the intervention of intermediaries. Then each ${\mathcal{W}}_j$ only needs to verify $\dfrac{1}{J}$ of the ZKPoKs during the execution. The checked ZKPoKs must be signed by the responsible workers. Theoretically, the total running time could be reduced by a factor of $J$. After the protocol, each worker can cross-check the ZKPoKs to ensure none of the workers fails to fulfil their duties without affecting the user experience. Once the dereliction of duty is found, any worker is encouraged to submit the invalid ZKPoKs to the smart contract for verification. Then the smart contract would check the ZKPoKs and the signatures before rewarding the rest of the workers with the deposit of the malicious worker.

\subsection{Generalisation}

Our median computation protocol can be generalised to the computation of the $k^{th}$ element. There are two changes to make:
\begin{itemize}
\item In step (1), the guessed $k^{th}$ value in the first round is initialised as $\lfloor (\overline{B}-\underline{B}) \cdot \dfrac{k}{N} + \underline{B} \rfloor + \dfrac{1}{2}$ and remains $\lfloor \dfrac{\alpha+\beta}{2} \rfloor + \dfrac{1}{2}$ for the subsequent rounds.

\item In step (2), the aggregation should be:
\begin{align*}
\begin{split}
z=&{\rm Dec}\big(\sum_{i=1}^{N, \oplus} \langle \phi(q_i)\rangle \oplus \langle 2k-N \rangle\big)\\
=&\sum_{i=1}^{N} \phi(q_i) + 2k-N 
\end{split}
\end{align*}
\end{itemize}
The protocol computes the median when $k=\dfrac{N}{2}$.

\section{Non-Interactive Protocol $\Pi_{\rm nirank}$} \label{sec:protocol2}

In this section, we present a non-interactive protocol $\Pi_{\rm nirank}$ to withstand the early-quit attack and mitigate the limitation of users having to remain online. The users only need to provide the inputs once and outsource the whole computation to the workers without involvement. Recall that the users are required to provide the encoded signs $\langle \phi\big(q_i=\eta \cdot (x_i-m)\big) \rangle$ in step (1) of $\Pi_{\rm irank}$, such that the workers can recursively search the target value. For the non-interactive protocol, the challenge consists in how the workers can help themselves to obtain the encoded sign $\langle \phi(q_i) \rangle$ without users' involvement while preserving users' privacy. Naively, the workers can decode each $\langle q_i \rangle$ as $q_i$ and re-encode $\phi(q_i)$ as $\langle \phi(q_i) \rangle$, which, nevertheless, leaks users' privacy. Thus, we propose a distributed masking scheme, which contributes to masking $q_i$ after the decode to preserve users' privacy.

\subsection{Distributed Masking Scheme}

In this section, we formalise the distributed masking scheme, which allows $J$ distributed parties to mask a non-zero value (either positive or negative) and its sign jointly. The scheme consists of a pair of PPT algorithms ({\rm Setup}, {\rm DM}). The {\rm Setup} algorithm generates a non-zero plaintext value space ${\mathcal{X}}$, a non-zero randomness space ${\mathcal{R}}$ and a non-zero masking space ${\mathcal{K}}$, where ${\mathcal{X}} \cap \{0\} =\emptyset,~{\mathcal{R}} \cap \{0\} =\emptyset,~{\mathcal{K}} \cap \{0\} =\emptyset$. The distributed masking algorithm {\rm DM} defines a function ${\mathcal{X}} \times {\mathcal{R}} \rightarrow {\mathcal{K}}$. To mask a non-zero value $x \in {\mathcal{X}}$, {\rm DM} can use a non-zero random value $r \in {\mathcal{R}}$ to generate a masking value $y=x \cdot r \in {\mathcal{K}}$, where
\begin{equation}
r=\prod_{j=1}^J r_j,~ r_j= (-1)^{\Xi_j} \prod_{\upsilon=1}^{\Upsilon_j} \tau_j^{(\upsilon)},~\tau_j^{(\upsilon)} \in {\mathcal{X}}
\end{equation}
$r_j$ is the product of a set of non-zero values $(\tau_j^{(\upsilon)})_{\upsilon=1}^{\Upsilon_j}$. $\Upsilon_j$ and $\Xi_j$ are two arbitrary positive integers produced by the $j$-th distributed party. 

The high-level idea is to enable a set of distributed parties to jointly generate a series of non-zero random values $\big((\tau_j^{(\upsilon)})_{\upsilon=1}^{\Upsilon_j}\big)_{j=1}^J$ to mask $x$, such that no adversaries can tell which is the target value $x$ and whether $x$ is positive or negative. Each $\tau_j^{(\upsilon)}$ must be in the same plaintext space as $x$ and $\Xi_j$ is a random number to mask the sign. Let each distinct value in the space ${\mathcal{X}}$ be a factor of the masking value $y$, and let the number of each factor be its frequency. Each distributed party is expected to sample randomly non-negative frequencies of all the factors from an arbitrary distribution that is only known to herself, e.g., uniform distribution, Gaussian distribution. This means that each party is allowed to miss a small portion of factors randomly. When all the random values $(r_j)_{j=1}^J$ and the target value $x$ are multiplied together, the factors generated by all the parties will complement each other, such that all the values in the space ${\mathcal{X}}$ would appear in $y$ with overwhelming probability. Allowing each party to miss of a portion of factors aims to generate the combined factors with arbitrary non-zero frequencies such that none of the parties can infer any privacy from the knowledge of the lowest frequency that can be generated. In fact, each party only needs to sample the frequencies of the random prime numbers in the plaintext space since the composite numbers consist of these prime numbers. 

Let $\chi_0$ be a variable representing the probabilistic distribution $x$ after seeing the masking value $y$ and $\chi_1$ be a variable indicating the uniform distribution of sample space ${\mathcal{X}}$. We formally define our scheme:

\smallskip
{\bf Theorem 6.1} {\em Assuming a group of $J$ distributed parties generating the randomness in the $(J, c)$-colluding setting, the distributed masking scheme (Setup, DM) is $(J-c)$-statistically hiding if the advantage of an adversary ${\mathcal{A}}$ in distinguishing $\chi_0$ and $\chi_1$ are statistically negligible in $(J-c)$, where $c$ is the number of corrupted parties:}
\begin{equation*}
\Big|Pr[\chi_0=\rho] - Pr[\chi_1=\rho] \Big|
= {\rm negl}(J-c)
\end{equation*}  
where ${\rm negl}(\cdot)$ is a negligible function and $\rho \in {\mathcal{X}}$ is an arbitrary value.

We disregard the random values generated by the corrupted parties since the security only relies on those $J-c$ honest, non-colluding parties who strictly follow the protocols. The adversary can create a factor frequency histogram by factorising $y$. Intuitively, if a value in ${\mathcal{X}}$ does not divide the masking value $y$, then the adversary would gain an advantage in distinguishing the two variables since the sample space of $\chi_0$ is reduced to $||{\mathcal{X}}||-1$ instead of $||{\mathcal{X}}||$. Otherwise, if all the values in ${\mathcal{X}}$ divide $y$, then it amounts to randomly guessing the target value $x$, which gives a negligible advantage to the adversary. Fig. \ref{fig:histogram} exemplifies a frequency histogram, where each factor has a non-zero frequency. From the histogram, it is intuitively infeasible to tell which is the target $x$.

\begin{figure}[!h]
\centering
\includegraphics[width=0.49\textwidth]{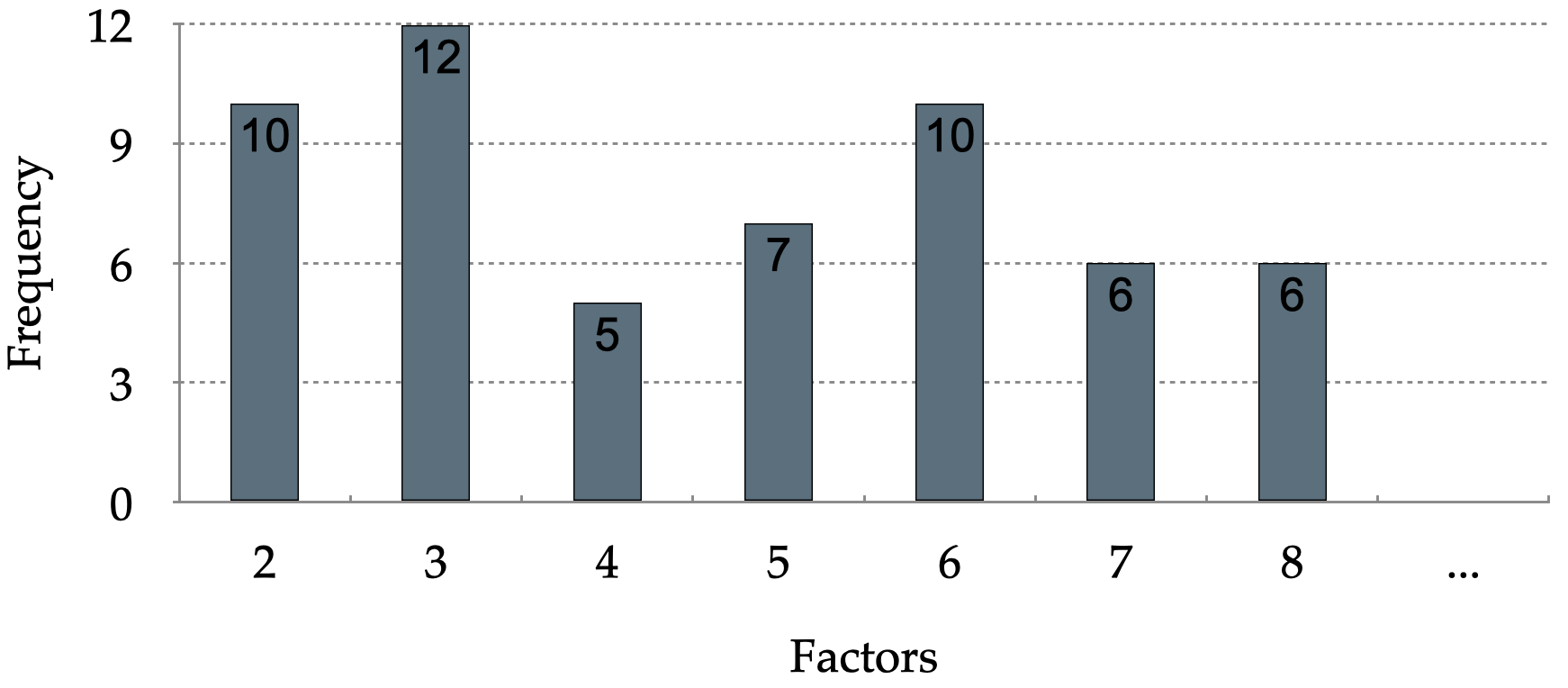}
\caption{An example of a frequency histogram.}
\label{fig:histogram}
\end{figure}

{\em Proof.} Distributed randomness is an effective technique to reduce the probability of producing zero-frequency factors and even low-frequency factors. When all the random values $(r_j)_{j=1}^{J-c}$ and $x$ are multiplied together to generate $y$, the frequencies of various factors would be added up. This would exponentially reduce the probability of producing zero-frequency factors. For example, assume each honest party samples random frequencies from a uniform distribution in an arbitrary frequency range ${\rm R_f^{(j)}}=\{0, 1, ...\}$ and let $p_j=\dfrac{1}{||{\rm R_f^{(j)}}||}$ be the probability of a factor having a certain frequency for the $j$-th party. Then the probability of a certain factor not showing up in the distributed random value $r$ would be $\prod_{j=1}^{J-c} p_j$. Based on the Bernoulli trial formula, the probability of producing at least one zero-frequency factor in $y$ would be $p=1-(1-\prod_{j=1}^{J-c} p_j)^{||{\mathcal{X}}||}$. Given a certain ${\mathcal{X}}$, $p$ would rapidly approach 0 when $J-c$ increases. More concretely, when $p_j=0.1$, $J-c=5$ and $||{\mathcal{X}}||=100$, $p$ would reach a negligible 0.099\%. $p$ would become substantially smaller if the honest parties reduce the probability of generating zero-frequency factors. Thus, it can be concluded that when $J-c$ is sufficiently large, the probability of producing low-frequency factors, such as those with zero-frequency or one-frequency, is also negligible.

For an adversary who only has $y$ and its factor frequency distribution but no clues to the honest parties' randomness, the probabilistic distribution of $\chi_0$ would be increasingly close to $\chi_1$ when $J-c$ grows larger, which gives it a negligible advantage in distinguishing the two variables. Note that the larger $||{\mathcal{X}}||$, the more honest parties are needed to achieve a negligible probability. Thus, in practice, we should use as many distributed parties as possible to achieve a negligible $p$ for a large plaintext space ${\mathcal{X}}$.

\subsection{Pre-processing Offline Phase} \label{sec:preprocessing}

The non-interactive protocol involves an additional {\em pre-processing offline phase}, where the workers ${\mathcal{W}}$ jointly execute a pre-processing protocol $\Pi_{\sf prep}$ to prepare a sufficient number of encoded random integers for the distributed masking scheme. Note that the offline phase is computationally expensive but can be conducted only once at any time before the online phase. ${\mathcal{W}}$ jointly generate:

\begin{enumerate}
\item a series of encoded random integers $\langle r \rangle=\prod_{j=1}^J \langle r_j \rangle$.

\item the encoded signs of the random integers $\langle \phi(r) \rangle=\prod_{j=1}^J \langle \phi(r_j) \rangle$.

\end{enumerate}
where $r_j,\phi(r_j)$ are generated by ${\mathcal{W}}_j$.

We formalise the language ${\mathcal{L}}_2$ of the ZKPoKs that each ${\mathcal{W}}_j$ must provide to demonstrate its compliance with the protocol:
$${\mathcal{L}}_2:=
\begin{cases}
{\rm stmt}=\big(\langle r_j \rangle, \langle \phi(r_j) \rangle, \langle \phi(r_j) \cdot r_j \rangle\big):\\
\exists {\rm wit}=\big(r_j, \phi(r_j)\big)~s.t.: \\
\phi(r_j) \in \{-1, 1\},~ \phi(r_j) \cdot r_j > 0\\
\end{cases}
$$
Both two constraints aim to ensure $\phi(r_j)$ and $r_j$ have the same sign and $r_j$ is non-zero.  Please refer to Appendix \ref{sec:append4} for the concrete instantiation.

\subsection{Online Phase} \label{sec:onlineprotocol}

In the online phase, for the $l$-th round, the workers ${\mathcal{W}}$ leveraging the distributed masking scheme to jointly obtain the encoded sign $\langle \phi(q_i) \rangle$ for each ${\mathcal{U}_i}$ without leaking the user's privacy, where $q_i=x_i-m_l$:
\begin{enumerate}
\item For each $\langle q_i \rangle$ in step (1), of $\Pi_{\sf irank}$, ${\mathcal{W}}$ jointly compute $\langle y_i \rangle$ with the encoded randomness $\langle r_i \rangle$ obtained in the pre-processing phase:
\begin{equation}
\langle y_i \rangle = \langle q_i \rangle \otimes \langle r_i \rangle=\langle q_i \cdot r_i \rangle
\label{eqn:mask1}
\end{equation}

\item ${\mathcal{W}}$ decode $\langle y_i \rangle$ to obtain $y_i=q_i \cdot r_i$, which statistically hides the value and sign of $q_i$.

\item${\mathcal{W}}$ extract its sign $\phi(y_i)$ and re-encodes it as $\langle \phi(y_i) \rangle$. 

\item ${\mathcal{W}}$ acquire the encoded sign of $q_i$:
\begin{equation}
\langle \phi(q_i) \rangle=\langle \phi(y_i) \rangle \otimes \langle \phi(r_i) \rangle = \langle \phi(q_i) \cdot \phi(r_i^2) \rangle
\label{eqn:mask2}
\end{equation}
where $\phi(r_i^2)=1$ for $r_i \neq 0$.

\end{enumerate}
Subsequently, the workers execute the step (2) of $\Pi_{\sf irank}$ by aggregating all the values $\langle \phi(q_i) \rangle$.

\section{$\Pi_{\sf irank}$ vs. $\Pi_{\sf nirank}$}

We compare the two protocols in Table \ref{tab:protocolcompare}. $\Pi_{\sf irank}$ requires the users to interact with the workers and share a portion of the computational workloads. It guarantees security even if at most $J-1$ workers collude together since the decode requires $J$ workers. However, $\Pi_{\sf irank}$ is vulnerable to the early-quit attack. $\Pi_{\sf nirank}$ removes users' interaction to withstand the early-quit attack by shifting the whole computational workload to the workers. Nevertheless, on the one hand, it requires an additional pre-processing phase. On the other hand, it relaxes the collusion tolerance to $c$ for using the distributed masking scheme, where $c$ depends on the size of the plaintext space. The larger the space, the more honest workers are needed for the distributed masking scheme. As is implied by ``No Free Lunch Theorem'' \cite{nofreelunch}, there is no single optimal algorithm for all kinds of scenarios. Both protocols have their strengths and weaknesses, which can be applied to different scenarios based on the actual requirements.

\begin{table}[!h]
\centering
\caption{Comparison of $\Pi_{\sf irank}$ with $\Pi_{\sf nirank}$}
\resizebox{0.45\textwidth}{!}{%
\begin{tabular}{@{}ccc@{}}
\toprule
Metric   & $\Pi_{\sf irank}$  & $\Pi_{\sf nirank}$ \\ \midrule
Interactivity     & \cmark & \xmark     \\ 
Pre-processing     & \xmark & \cmark     \\ 
Collusion tolerance of workers         &   $J-1$  &  Depends on $||{\mathcal{X}}||$  \\
\bottomrule
\end{tabular}%
}
\label{tab:protocolcompare}
\end{table}

\begin{table}[!h]
\centering
\caption{Comparison of workers' computational workloads in each round, where $N$ is the dataset size.}
\resizebox{0.28\textwidth}{!}{%
\begin{tabular}{@{}ccccc@{}}
\toprule
\textbf{Type} & ${\rm Enc}$ & ${\rm Dec}$ & $\otimes$ & $\oplus$ \\ \midrule
   $\Pi_{\sf irank}$  &   0 & 1 & 0 & N-1 \\
   $\Pi_{\sf nirank}$ &  N & N+1 & 2N & N-1 \\
\bottomrule
\end{tabular}
}
\label{tab:workercomparison}
\end{table}

Table \ref{tab:workercomparison} shows a concrete comparison of the workers' computational workloads in a single round of the two protocols. $\Pi_{\sf irank}$ does not demand many computational resources of the workers, who only need to perform $N-1$ additive operations on encoded values and one decode operation. In contrast, for $\Pi_{\sf nirank}$, the workers need to perform 2N multiplicative and N-1 additive operations on encoded values as well as N encode and N+1 decode operations. In summary, our protocols only entail $O(N)$ time complexity for a single round.

\section{Union of Multiple Datasets}

In this scenario, we consider $N$ users, where each user $N$ holds a dataset $D_i$ and all the elements lie in a public range $[\alpha, \beta]$. For our computing model, Scenario-I is a special case of Scenario-II, where $D_i$ has only one element. The users need to encode all the elements and jointly run the protocols to compute the global $k^{th}$ element of all the datasets without leaking the privacy of the local distribution or individual data. The time complexity would be $O(\sum_{i=1}^N ||D_i||)$.

\section{Instantiation \& Proofs of Security} \label{sec:instantiation}

We provide complete and self-contained construction of our computing model by instantiating the Enc-Dec scheme with the threshold Paillier cryptosystem. We choose Paillier over elliptic-curve Elgamal\footnote{ECC Elgamal is used by the other two HE-based solutions \cite{blockchainmedian, starmedian}.} as the latter suffers from a limitation of only supporting small integers. We employ a tuple of Paillier-based $\Sigma$-protocol zero-knowledge proofs to defend against static, active (malicious) adversaries in the UC framework. Note that our computing model can also be instantiated with somewhat homomorphic encryptions (SHE) \cite{SHE} or secret-sharing schemes, e.g., Commitment-Enhanced SPDZ framework \cite{cespdz}.

\subsection{Enc-Dec Scheme Instantiation}

\subsubsection{Paillier Cryptosystem}

The Paillier cryptosystem \cite{paillier} is a popular additively homomorphic cryptosystem based on the decisional composite residuosity assumption (DCRA). It allows to perform computations on the encrypted data and output the results as if the operations are performed on unencrypted data (plaintext). It encrypts a plaintext $\theta \in {\mathbb{Z}}_n$ as a ciphertext ${\mathcal{E}}(\theta) \in {\mathbb{Z}}_{n^2}$:
$${\mathcal{E}}(\theta)=g^{\theta} \cdot r^n \  (\mbox{mod\ }n^{2})$$
where ${\mathbb{Z}}_n$ is an RSA cyclic group of an unknown prime order $n=p \cdot q$. $p$ and $q$ are large random prime integers. $g$ is a generator and $r \in {\mathbb {Z}}_n^*$ is a random integer. ${\mathbb {Z}}_n^*={\mathbb {Z}}_n\backslash\{0\}$.

\noindent
The Paillier cryptosystem has two homomorphic properties:

\begin{itemize}
\item The product of two ciphertexts ${\mathcal E}(\theta_1)$ and ${\mathcal E}(\theta_2)$ can be decrypted as the sum of two plaintexts $\theta_1+\theta_2$:
\begin{equation}
    {\mathcal D}\big({\mathcal E}(\theta_1) \cdot {\mathcal E}(\theta_2) \  (\mbox{mod\ }n^{2}) \big) = \theta_1+\theta_2 \ (\mbox{mod\ } n) \label{eqn:paillier-add}
\end{equation}

\item A ciphertext ${\mathcal E}(\theta_1)$ raised to the power of $\theta_2$ can be decrypted as the product of the two plaintexts $\theta_1 \cdot \theta_2$:
\begin{equation}
   {\mathcal D}\big({\mathcal E}(\theta_1)^{\theta_2}  \  (\mbox{mod\ }n^{2})\big) = \theta_1 \cdot \theta_2   \ (\mbox{mod\ } n) \label{eqn:paillier-mul}
\end{equation}
\end{itemize}
where ${\mathcal E}(\cdot)$ and ${\mathcal D}(\cdot)$ indicate the encryption and decryption operations, which implement the Enc() and Dec() algorithms, respectively. Note that we will omit $(\text{mod}~n^{2})$ and $(\text{mod}~n)$ in the following paper for brevity.

Furthermore, to support negative numbers, we shift the allowed range $[0, n-1]$ to $[-\frac{n-1}{2}, \frac{n-1}{2}]$. For instance, when a decrypted value $\theta$ is greater than $\frac{n-1}{2}$, we shift back $\theta$ to the range $[-\frac{n-1}{2}, \frac{n-1}{2}]$ by computing $\theta-n$.

\subsubsection{Our Instantiation}

We adopt an $(N, t)$-threshold variant of the Paillier cryptosystem, namely, the well-known Damg{\aa}rd-Jurik cryptosystem \cite{thresholdpaillier}. This variant uses one public key for encryption and multiple private keys for decryption. Each private key can be used to produce a "partial decryption share" of a ciphertext. One can combine more than $t$ partial shares to obtain the original plaintext. The variant is equipped with a $\Sigma$-protocol zero-knowledge proof of partial decryption (zkpPD) to prove a partial decryption share is consistent with the corresponding private key. We allow each worker server to hold a private key to perform decryptions jointly.

The two homomorphic properties in Eqn. \eqref{eqn:paillier-add} and \eqref{eqn:paillier-mul} satisfy the encode operations $\langle \theta \rangle \oplus \langle \delta \rangle=\langle \theta+\delta \rangle$ and $\langle \theta \rangle \otimes \delta =\langle \theta \cdot \delta \rangle$. To implement the multiplicative operation $\langle \theta \rangle \otimes \langle \delta \rangle=\langle \theta \cdot \delta \rangle$ with the Paillier cryptosystem, we need to make a slight transformation:
$$\langle \theta \rangle \otimes \langle \delta \rangle = \sum_{j=1}^{J, \oplus} (\langle \theta \rangle \otimes \langle \delta_j \rangle)= \langle \theta \cdot \sum_{j=1}^J \delta_j \rangle =\langle \theta \cdot \delta \rangle,~\delta=\sum_{j=1}^J \delta_j$$
Concretely, we employ a secret-sharing protocol $\Pi_{\sf sr}$ to secretly share a ciphertext ${\mathcal E}(\delta)$ among $J$ parties so that the $j$-th party holds a share ${\mathcal E}(\delta_j)$. Then, the $j$-th party broadcasts a zkpMTP$\{{\mathcal E}(\theta), {\mathcal E}(\delta_i), {\mathcal E}( \theta \cdot \delta_j)\}$. Finally, all the parties perform the aggregation $\prod_{j=1}^J {\mathcal E}(\theta\cdot \delta_j)={\mathcal E}(\theta \cdot \sum_{j=1}^J \delta_j)={\mathcal E}(\theta \cdot \delta)$. We will use this technique to perform the multiplicative operations on encrypted values in Eqn. \eqref{eqn:mask1} and \eqref{eqn:mask2}. Note that the encrypted randomness prepared in the pre-processing phase of $\Pi_{\sf nirank}$ must be secretly shared among the workers before the online phase. Please see the Paillier-based sub-protocols $\Pi_{\sf sr}$ and $\Pi_{\sf prep}$ in Appendix \ref{sec:append3} and \ref{sec:append4}.

In our instantiation, we use the following settings for the distributed masking scheme:
\begin{itemize}
\item The plaintext space ${\mathcal{X}}$ is $R=[\underline{B}, \overline{B}] \cap \{0\}=\emptyset$.

\item The randomness space ${\mathcal{R}}$ is set to $[-\frac{n-1}{16(\overline{B}-\underline{B})}, 0) \cup (0,\frac{n-1}{16(\overline{B}-\underline{B})}]$ to ensure the masking space ${\mathcal{K}} \subseteq [-\frac{n-1}{2}, \frac{n-1}{2}]$, where $n$ is exponentially larger than $||R||=(\overline{B}-\underline{B})$.

\end{itemize}

\subsection{ZKPoK Instantiation} \label{sec:zkpinstantiation}

Firstly, for $\Pi_{\sf irank}$, we describe the Paillier-based ZKPoKs to satisfy the relations in the language ${\mathcal{L}}_1$. Please refer to Appendix \ref{sec:append1} for the concrete protocols of ZKPoKs.

\begin{enumerate}

\item Constraint (1) requires:
\begin{itemize}
\item a zkpMBS$\big\{{\mathcal{E}}\big(\phi(q_i)\big), \{-1, 1\}\big\}$ 
\end{itemize}

\item Constraint (2) requires:
\begin{itemize}

\item a zkpMTP$\{{\mathcal{E}}\big(\phi(q_i)\big), {\mathcal{E}}(q_i), {\mathcal{E}}\big(\phi(q_i) \cdot q_i\big)\}$ 

\item a zkpRG$\{{\mathcal{E}}\big(\phi(q_i) \cdot q_i\big), [0, \overline{B}] \}$

\item a zkpNZ$\{{\mathcal{E}}\big(\phi(q_i) \cdot q_i\big), \phi(q_i) \cdot q_i \neq 0\}$

\end{itemize}
\end{enumerate}
Note that one can naively use two zkpRGs to prove $\phi(q_i) \cdot q_i \in [1, \overline{B}]$. Instead, in our instantiation, we use one zkpRG and one zkpNZ to prove $\phi(q_i) \cdot q_i \in (0, \overline{B}]$ as a zkpNZ is more efficient than a zkpRG (Please refer to Section \ref{sec:experiment} for efficiency comparisons). 

Secondly, for $\Pi_{\sf nirank}$ in Section \ref{sec:onlineprotocol}, apart from the zkpPDs, the workers must use two zkpMTPs to perform the computations in Eqn. \eqref{eqn:mask1} and \eqref{eqn:mask2} for each user in one single round.

The Paillier-based zkpMTP$\{{\mathcal{E}}(\theta), {\mathcal{E}}(\delta), {\mathcal{E}}(\rho)\}$ only proves the relation $\rho=\theta \cdot \delta~(\text{mod}~n)$ within the finite field ${\mathbb{Z}}_n$. Thus, we will use zkpRGs or zkpMBSs to confine the range of $\theta$ and $\delta$ so as to ensure $\rho$ is within $[-\frac{n-1}{2}, \frac{n-1}{2}]$. In our protocol, each ${\mathcal{U}}_i$ is required to provide a zkpRG for the input $x_i$ to ensure the corresponding $q_i=x_i-m$ lies in the allowed range.

\subsection{Proofs of Security}

\subsubsection{Universal Composability (UC)} \label{sec:ucmodel}

The UC-framework \cite{ucmodel} guarantees stronger security than the stand-alone model. It allows a protocol $\Pi$ to run concurrently with any arbitrary protocols in a secure manner. In the UC framework, there is an adversarial {\em environment} ${\mathcal{Z}}$ and a {\em dummy adversary} ${\mathcal{A}}$ who relays messages between the corrupted parties and ${\mathcal{Z}}$. The UC framework defines a {\em real} world and an {\em ideal} world, where ${\mathcal{Z}}$ provides inputs for all the parties and receives their outputs. Besides, ${\mathcal{Z}}$ can corrupt parties and take control over their actions. In the real world, the parties and the real adversary jointly run the protocol $\Pi$. In the ideal world, there exists a {\em simulator} ${\mathcal{S}}$ interacting with ${\mathcal{Z}}$ and an {\em ideal functionality} ${\mathcal{F}}$. The ideal functionality honestly performs the prescribed computations of the protocol $\Pi$ and returns the correct outputs. ${\mathcal{S}}$ has to simulate an indistinguishable view from $\Pi$ without rewinding to confuse ${\mathcal{Z}}$ about the two worlds. We say $\Pi$ UC-emulates ${\mathcal{F}}$ if ${\mathcal{Z}}$ cannot distinguish the real world and the ideal world. In addition, UC-framework offers modular composition security. If another protocol, $\Omega$, uses $\Pi$ as a sub-protocol, the calls to $\Pi$ can be replaced with the calls to ${\mathcal{F}}$. We can say $\Omega^{\Pi}$ UC-emulates $\Omega^{\mathcal{F}}$.

\subsubsection{Trusted Setup}

As is stated in the SPDZ paper \cite{spdz}, some setup assumption is required to demonstrate UC security. In light of this, we assume an ideal functionality ${\mathcal{F}}_{\sf keygen}$ that generates and distributes a key pair among the parties for the threshold Paillier cryptosystem. Some UC-secure MPC protocols such as \cite{secretshareprotocol}, can be used to UC-emulate ${\mathcal{F}}_{\sf keygen}$. Let $H$ and $C$ be a group of honest and corrupted parties.

\medskip
\begin{longfbox}[border-break-style=none,border-color=\#bbbbbb,background-color=\#eeeeee,breakable=true,,width=\linewidth]
\hspace*{26mm} Functionality ${\mathcal{F}}_{\sf keygen}$

\smallskip
1. On input ${\bf start}$, run the algorithm $({\sf pk}, {\sf sk}) \leftarrow {\sf KeyGen}(\lambda)$ to generate a public encryption key ${\sf pk}$ and a secret decryption key ${\sf sk}$.

\smallskip
3. On input ${\bf receive}$, receive a set of secret shares ${\sf sk_j}$ from ${\mathcal{A}}$ for ${\mathcal{W}}_j \in C$ and send the secret shares ${\sf sk_j}$ for ${\mathcal{W}}_j \in H$. Send ${\sf pk}$ to all and store $({\sf sk_j})_{j=1}^J$.

\end{longfbox}
\begin{center}Figure 2: The ideal functionality ${\mathcal{F}}_{\sf keygen}$\end{center}

\subsubsection{Other Ideal Functionalities} 

Before describing the proofs of security, we first define a series of ideal functionalities that run with the users $({\mathcal{U}}_i)_{i=1}^N$, the workers $({\mathcal{W}}_j)_{j=1}^J$ and a simulator ${\mathcal{S}}$. 
\begin{itemize}
\item{${\mathcal{F}}_{\sf ddec}$:} It handles the distributed decryption.

\item{${\mathcal{F}}_{\sf prep}$:} It generates a sufficient number of random integers for the preprocessing phase of $\Pi_{\sf nirank}$.

\item{${\mathcal{F}}_{\sf irank}$ \& ${\mathcal{F}}_{\sf nirank}$:} They deal with all the procedures of a single round of the two protocols. 

\end{itemize}

\medskip
\begin{longfbox}[border-break-style=none,border-color=\#bbbbbb,background-color=\#eeeeee,breakable=true,,width=\linewidth]
\hspace*{25mm} Functionality ${\mathcal{F}}_{\sf ddec}$

\smallskip
1. On input $\big({\bf decrypt}, {\mathcal{E}}(z)\big)$, send $z$ to all.

\smallskip
2. On input ${\bf abort}$, send abort to all.

\end{longfbox}
\begin{center}Figure 3: The ideal functionality ${\mathcal{F}}_{\sf ddec}$\end{center}

\smallskip

\begin{longfbox}[border-break-style=none,border-color=\#bbbbbb,background-color=\#eeeeee,breakable=true,,width=\linewidth]
\hspace*{25mm} Functionality ${\mathcal{F}}_{\sf prep}$

\smallskip
1. On input ${\bf receive}$, receive a set of random integers $\big(r_j, \phi(r_j)\big)$ for ${\mathcal{W}}_j \in C$ from ${\mathcal{A}}$ and send a set of random integers to ${\mathcal{W}}_j \in H$.

\smallskip
2. On input ${\bf abort}$, send abort to all.

\end{longfbox}
\begin{center}Figure 4: The ideal functionality ${\mathcal{F}}_{\sf prep}$\end{center}

\smallskip

\begin{longfbox}[border-break-style=none,border-color=\#bbbbbb,background-color=\#eeeeee,breakable=true,,width=\linewidth]
\hspace*{25mm} Functionality ${\mathcal{F}}_{\sf irank}$

\smallskip
1. On input $\big({\bf input}, {\mathcal{U}}_i, x_i, \phi(x_i-m_l)\big)$, store it.

\smallskip
2. On input ${\bf output}$, compute the result $z$ and send it to ${\mathcal{S}}$.

\smallskip
3. On input ${\bf abort}$, send abort to all.

\end{longfbox}
\begin{center}Figure 5: The ideal functionality ${\mathcal{F}}_{\sf irank}$ of the $l$-th round\end{center}

\smallskip

\begin{longfbox}[border-break-style=none,border-color=\#bbbbbb,background-color=\#eeeeee,breakable=true,,width=\linewidth]
\hspace*{25mm} Functionality ${\mathcal{F}}_{\sf nirank}$

\smallskip
1. On input $\big({\bf input}, {\mathcal{U}}_i, x_i\big)$, store it.

\smallskip
2. On input ${\bf output}$, compute the result $z$ and send it to ${\mathcal{S}}$.

\smallskip
3. On input ${\bf abort}$, send abort to all.

\end{longfbox}
\begin{center}Figure 6: The ideal functionality ${\mathcal{F}}_{\sf nirank}$ of the $l$-th round\end{center}
\smallskip

\noindent
We provide two theorems as below and describe the two protocols $\Pi_{\sf ddec}$ and $\Pi_{\sf prep}$ in Appendix \ref{sec:append2} and \ref{sec:append4}.

\medskip
{\bf Theorem 1.} {\em Assuming DCRA is hard and the NIZKPoKs are secure in the random oracle, our threshold Paillier-based interactive protocol hybrid $\Pi_{\sf ddec}^{\mathcal{F}_{\sf keygen}}$ securely UC-emulates ${\mathcal{F}}_{\sf ddec}$ with identifiable abort with respect to a dummy adversary ${\mathcal{A}}$ if there exists a PPT simulator ${\mathcal{S}}$ such that for every PPT, static, active environment ${\mathcal{Z}}$ corrupting no more than $J-1$ workers, it holds that:}
$$\{{\rm IDEAL}_{{\mathcal{F}}_{\sf ddec}, {\mathcal{S}}}^{{\mathcal{F}}_{\sf keygen}}\} \overset{comp}{\equiv} \{{\rm REAL}_{\Pi_{\sf ddec}, {\mathcal{A}}}^{{\mathcal{F}}_{\sf keygen}}\}$$
where {\rm IDEAL} and {\rm REAL} respectively refer to the views of ${\mathcal{Z}}$ that consist of the views and outputs of the corrupted and honest parties in the ideal and real world. $\overset{comp}{\equiv}$ means two views are computationally indistinguishable.

\medskip
{\bf Theorem 2.} {\em Assuming DCRA is hard and the NIZKPoKs are secure in the random oracle, our threshold Paillier-based interactive protocol hybrid $\Pi_{\sf prep}^{{\mathcal{F}}_{\sf keygen}, {\mathcal{F}}_{\sf ddec}}$ securely UC-emulates ${\mathcal{F}}_{\sf prep}$ with identifiable abort with respect to a dummy adversary ${\mathcal{A}}$ if there exists a PPT simulator ${\mathcal{S}}$ such that for every PPT, static, active environment ${\mathcal{Z}}$ corrupting no more than $J-1$ workers, it holds that:}
$$\{{\rm IDEAL}_{{\mathcal{F}}_{\sf prep}, {\mathcal{S}}}^{{\mathcal{F}}_{\sf keygen}, {\mathcal{F}}_{\sf ddec}}\} \overset{comp}{\equiv} \{{\rm REAL}_{\Pi_{\sf prep}, {\mathcal{A}}}^{{\mathcal{F}}_{\sf keygen}, {\mathcal{F}}_{\sf ddec}}\}$$

\subsubsection{Proofs}

We provide the security proofs for our instantiated threshold Paillier-based interactive protocol $\Pi_{\sf irank}$ and non-interactive protocol $\Pi_{\sf nirank}$. We set $t=J$ to require all the workers to perform the decryption.

\medskip
{\bf Theorem 3.} {\em Assuming DCRA is hard and the NIZKPoKs are secure in the random oracle, our threshold Paillier-based interactive protocol hybrid $\Pi_{\sf irank}^{\mathcal{F}_{\sf keygen}, {\mathcal{F}}_{\sf ddec}}$ securely UC-emulates ${\mathcal{F}}_{\sf irank}$ with identifiable abort with respect to a dummy adversary ${\mathcal{A}}$ if there exists a PPT simulator ${\mathcal{S}}$ such that for every PPT, static, active environment ${\mathcal{Z}}$ corrupting no more than $J-1$ workers, it holds that:}
$$\{{\rm IDEAL}_{{\mathcal{F}}_{\sf irank}, {\mathcal{S}}}^{{\mathcal{F}}_{\sf keygen}, {\mathcal{F}}_{\sf ddec}}\} \overset{comp}{\equiv} \{{\rm REAL}_{\Pi_{\sf irank}, {\mathcal{A}}}^{{\mathcal{F}}_{\sf keygen}, {\mathcal{F}}_{\sf ddec}}\}$$

We construct a straight-line UC-simulator ${\mathcal{S}}$ that internally runs copies of the real protocol $\Pi_{\sf irank}$, and externally interacts with a non-rewindable environment ${\mathcal{Z}}$ and the ideal functionality ${\mathcal{F}_{\sf irank}}$. ${\mathcal{S}}$ can directly extract the corrupted parties' inputs without rewinding since it learns the decryption keys of the cryptosystem from ${\mathcal{F}_{\sf keygen}}$. Besides, we adopted the simulation technique \cite{grothvoting} for threshold encryption cryptosystems that ${\mathcal{S}}$ can control the random oracle ${\mathcal{O}}$ to simulate NIZKPoKs. Concretely, ${\mathcal{S}}$ is empowered to pre-assign a response to a query to ${\mathcal{O}}$. With this capability, ${\mathcal{S}}$ can simulate the partial decryption shares of a ciphertext and the corresponding NIZKPoKs for honest workers, such that the ciphertext can decrypt to the desired value.

\medskip
\begin{longfbox}[border-break-style=none,border-color=\#bbbbbb,background-color=\#eeeeee,breakable=true,,width=\linewidth]
\hspace*{29mm} Simulator ${\mathcal{S}_{\sf irank}}$

\smallskip
1. On input {\bf initialise}, ${\mathcal{S}_{\sf irank}}$ calls ${\mathcal{F}_{\sf keygen}}$ to generate a key pair. Then it knows all the secret decryption keys to decrypt the encrypted messages of the corrupted parties.

\smallskip
2. On input {\bf input}, for ${\mathcal{U}}_i \in H$, ${\mathcal{S}_{\sf irank}}$ provides dummy encrypted inputs ${\mathcal{E}}(x_i)$ and ${\mathcal{E}}\big(\phi(x_i-m_l)\big)$ for an arbitrary $x_i \in {\mathcal{X}}$ with simulated NIZKPoKs in each round. For ${\mathcal{U}}_i \in C$, ${\mathcal{Z}}$ provides the encrypted inputs along with the NIZKPoKs. 

\smallskip
3. On input {\bf output}, ${\mathcal{S}_{\sf irank}}$ acquires the actual output $z$ from ${\mathcal{F}}_{\sf irank}$. Then, ${\mathcal{S}_{\sf irank}}$ simulates the partial decryption share of an honest worker by controlling the random oracle, such that the aggregated result decrypts to $z$. ${\mathcal{S}_{\sf irank}}$ also simulates the corresponding zkpPD to validate the share.

\smallskip
4. On input {\bf abort}, ${\mathcal{S}_{\sf irank}}$ sends abort to ${\mathcal{F}_{\sf irank}}$ if any NIZKPoK provided by ${\mathcal{Z}}$ is invalid.

\end{longfbox}
\begin{center}Figure 7: The simulator ${\mathcal{S}_{\sf irank}}$ for $\Pi_{\sf irank}$ \end{center}

Note that ${\mathcal{S}_{\sf irank}}$ makes use of the zero-knowledge property of the NIZKPoKs and the computationally hiding property of the threshold Paillier cryptosystem to produce a computationally indistinguishable view from the real protocol.

\medskip
{\bf Theorem 4.} {\em Assuming DCRA is hard and the NIZKPoKs are secure in the random oracle, our threshold Paillier-based non-interactive protocol hybrid $\Pi_{\sf nirank}^{{\mathcal{F}_{\sf keygen}}, {\mathcal{F}}_{\sf ddec}, {\mathcal{F}_{\sf prep}}}$ securely UC-emulates ${\mathcal{F}}_{\sf nirank}$ with identifiable abort with respect to a dummy adversary ${\mathcal{A}}$ if there exists a PPT simulator ${\mathcal{S}}$ such that for every PPT, static, active  environment ${\mathcal{Z}}$ corrupting no more than $c$ workers, it holds that:}
$$\{{\rm IDEAL}_{{\mathcal{F}}_{\sf nirank}, {\mathcal{S}}}^{{\mathcal{F}}_{\sf keygen}, {\mathcal{F}}_{\sf ddec}, {\mathcal{F}_{\sf prep}}}\} \overset{comp}{\equiv} \{{\rm REAL}_{\Pi_{\sf nirank}, {\mathcal{A}}}^{{\mathcal{F}}_{\sf keygen}, {\mathcal{F}}_{\sf ddec}, {\mathcal{F}_{\sf prep}}}\}$$
\medskip
\begin{longfbox}[border-break-style=none,border-color=\#bbbbbb,background-color=\#eeeeee,breakable=true,,width=\linewidth]
\hspace*{29mm} Simulator ${\mathcal{S}_{\sf nirank}}$

\smallskip
1. On input {\bf initialise}, ${\mathcal{S}_{\sf irank}}$ calls ${\mathcal{F}_{\sf keygen}}$ to generate a key pair. ${\mathcal{S}_{\sf irank}}$ also calls ${\mathcal{F}}_{\sf prep}$ to create a sufficient number of encrypted randomness for the distributed masking scheme. ${\mathcal{S}_{\sf nirank}}$ knows the random values of the corrupted parties since it has decryption keys.

\smallskip
2. On input {\bf input}, ${\mathcal{S}_{\sf nirank}}$ does the same as ${\mathcal{S}_{\sf irank}}$.

\smallskip
3. On input {\bf mask}, ${\mathcal{S}_{\sf nirank}}$ follows the distributed masking scheme to simulate random values and their ciphertexts for the honest parties. ${\mathcal{S}_{\sf nirank}}$ uses the corrupted parties' random values and the simulated random values to generate a masking value $y_i$ for each ${\mathcal{U}}_i$. ${\mathcal{S}_{\sf nirank}}$ also simulates the 
the partial shares and the NIZKPoKs of the honest parties for the distributed masking process. 

\smallskip
3. On input {\bf output}, ${\mathcal{S}_{\sf nirank}}$ does the same as ${\mathcal{S}_{\sf irank}}$.

\smallskip
4. On input {\bf abort}, ${\mathcal{S}_{\sf nirank}}$ does the same as ${\mathcal{S}_{\sf irank}}$.

\end{longfbox}
\begin{center}Figure 8: The simulator ${\mathcal{S}_{\sf nirank}}$ for $\Pi_{\sf nirank}$ \end{center}

Note that without knowing honest workers' randomness, the PPT environment ${\mathcal{Z}}$ is unable to distinguish which value is masked by $y$ due to the hiding property of the distributed masking scheme.

\section{Empirical Experiments} \label{sec:experiment}

We evaluated the performance of our Paillier-based instantiation with respect to the accuracy, communication overhead and computational overhead. We adopted the Paillier Threshold Encryption Toolbox \cite{toolbox} for implementation. Our empirical evaluation was conducted with the processor Intel Core i7-8700 CPU @3.2GHz. All the experiments were executed on Java Virtual Machine in a single thread, with results averaged over 2,000 instances.

We used a 2048-bit RSA modulus $n$ for the Paillier cryptosystem, which guarantees 112-bit security and is assumed to be secure before 2030 based on the NIST recommendations \cite{keylength}. For the ease of measurement, we respectively used 2048-bit (256 bytes) and 4096-bit (512 bytes) to measure the size of the elements in ${\mathbb{Z}}_n$ and ${\mathbb{Z}}_{n^2}$. Besides, we used an 8-bit range size $[0, 127]$ for the range proofs. Note that smaller moduli, e.g., 1536-bit, 1792-bit, can also be used to pursue higher efficiency as the latest record of breaking RSA keys is 829-bit in 2020\footnote{https://phys.org/news/2020-03-cryptographic.html}, to the best of our knowledge. We assumed the worker servers possess adequate computational and communication resources. The users only communicate with the geographically closest server, so the communication network latency is relatively low.

\subsection{Accuracy}

We conducted an accuracy simulation of computing the $25^{th}$, $50^{th}$ and $75^{th}$ percentiles over different Gaussian distributions with $N=10,001$, $\mu=100$ and a series of standard deviations $\sigma$ based on the techniques of our protocols. We used the mean absolute errors ${\rm MAE}=\dfrac{1}{N} \sum_{i=1}^N |\xi-\hat{\xi}|$ to evaluate the accuracy, where $\xi$ and $\hat{\xi}$ indicate the actual and the expected results. In Figure \ref{fig:accuracy}, it can be observed that the MAEs fluctuate slightly between the negligible errors 0.5 and 0.7, which corroborates the high accuracy that our protocols can achieve.

\begin{figure}[!h]
\setcounter{figure}{7}
\centering
\includegraphics[width=0.41\textwidth]{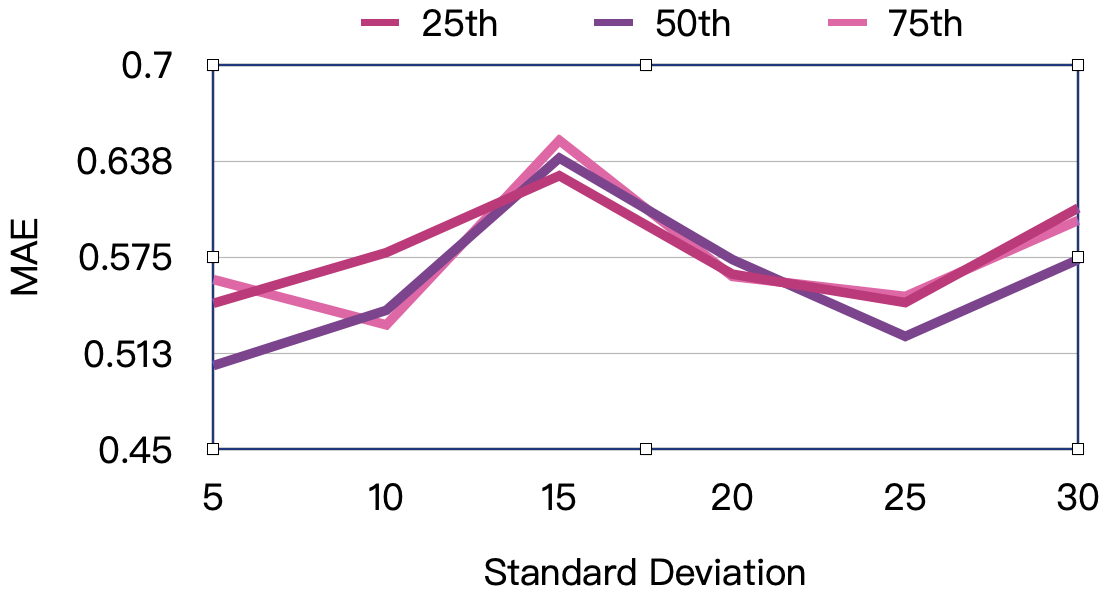}
\caption{Accuracy simulation.}
\label{fig:accuracy}
\end{figure}

\begin{table*}[!t]
\centering
\caption{The communication costs in kilobytes of zero-knowledge proofs, where the zkpMBS is with the witness $x \in \{-1, 1\}$ and zkpPD indicates the zero-knowledge proofs partial decryption.}
\resizebox{0.55\textwidth}{!}{%
\begin{tabular}{@{}cccccc@{}}
\toprule
\multicolumn{1}{l}{\textbf{Type}} & \multicolumn{1}{l}{\textbf{zkpMTP}} & \multicolumn{1}{l}{\textbf{zkpMBS}} & \multicolumn{1}{l}{\textbf{zkpRG}} & \multicolumn{1}{l}{\textbf{zkpNZ}} & \multicolumn{1}{l}{\textbf{zkpPD}}\\ \midrule
   No. of ${\mathbb{Z}}_n$                               & 1                                   & 1                                   & 5                                  & 1             & 1                     \\
   No. of ${\mathbb{Z}}_{n^2}$                                & 4                                   & 5                                   & 9                                 & 5                 & 2                 \\
KB                            & 2.25                                & 2.75                                & 5.75                               & 2.75          & 1.25                     \\ \bottomrule
\end{tabular}
}
\label{tab:zkpcommunicationcost}
\end{table*}
\begin{table*}[!t]
\centering
\caption{The average outbound communication cost comparison of two protocols in kilobytes, where $N$ and $R$ indicate the number of users and the input range.}
\resizebox{0.55\textwidth}{!}{%
\begin{tabular}{@{}ccc@{}}
\toprule
\textbf{Type} & \textbf{$\Pi_{\sf irank}$} & \textbf{$\Pi_{\sf nirank}$} \\ \midrule
   User (KB)                              & $6.25+15 \cdot (\lceil \log ||R|| \rceil-1)$                                   & 6.25 \\
   Server (KB)                               & $1.5 \cdot (\lceil \log ||R|| \rceil-1)$                                   & $5.75N \cdot (\lceil \log ||R|| \rceil-1)$  \\
\bottomrule
\end{tabular}
}
\label{tab:outcommunicationcomparison}
\end{table*}
\begin{table*}[!t]
\centering
\caption{The average inbound communication cost comparison of two protocols in kilobytes.}
\resizebox{0.55\textwidth}{!}{%
\begin{tabular}{@{}ccc@{}}
\toprule
                            & \textbf{Type}  &\multicolumn{1}{c}{\textbf{Communication Cost (KB)}} \\ \midrule
\multicolumn{1}{c}{\multirow{2}{*}{User}} & \textbf{$\Pi_{\sf irank}$}        &      $0.25 \cdot (\lceil \log ||R|| \rceil-1)$                                                                                                                             \\
\multicolumn{1}{c}{}                      & \textbf{$\Pi_{\sf nirank}$}       &       0.25                     \\
\multirow{2}{*}{Server}                   & \textbf{$\Pi_{\sf irank}$}        &      $\dfrac{N}{J} \cdot \big(6.25+15 \cdot (\lceil \log ||R|| \rceil-1)\big)$                                                                      \\
                                          & \textbf{$\Pi_{\sf nirank}$}       &   $\dfrac{1}{J} \cdot \big(6.25 N + 5.75N \cdot (J-1) \cdot (\lceil \log ||R|| \rceil-1)\big)$                      \\ \bottomrule 
\end{tabular}
}
\label{tab:incommunicationcomparison}
\end{table*}
\begin{table*}[!t]
\centering
\caption{The running time in milliseconds of zero-knowledge proofs, where the zkpMBS is with the witness $x \in \{-1, 1\}$.}
\resizebox{0.56\textwidth}{!}{%
\begin{tabular}{@{}cccccc@{}}
\toprule
\multicolumn{1}{l}{\textbf{Type}} & \multicolumn{1}{l}{\textbf{zkpMTP}} & \multicolumn{1}{l}{\textbf{zkpMBS}} & \multicolumn{1}{l}{\textbf{zkpRG}} & \multicolumn{1}{l}{\textbf{zkpNZ}} & \multicolumn{1}{l}{\textbf{zkpPD}}\\ \midrule
   Prover (ms)                              & 73                                   & 93                                   & 148                                  & 99             & 91                     \\
   Verifier (ms)                               & 49                                   & 60                         & 99                                 & 58                 & 71                 \\
 \bottomrule
\end{tabular}
}
\label{tab:zkpcomputationalcost}
\end{table*}
\begin{table*}[!t]
\centering
\caption{The computational cost comparison in milliseconds.}
\resizebox{0.65\textwidth}{!}{%
\begin{tabular}{@{}ccc@{}}
\toprule
                            & \textbf{Type}  &\multicolumn{1}{c}{\textbf{Running Time (ms)}} \\ \midrule
\multicolumn{1}{c}{\multirow{2}{*}{User}} & \textbf{$\Pi_{\sf irank}$}        &      $159+413 \cdot (\lceil \log ||R|| \rceil-1)$                                                        \\
\multicolumn{1}{c}{}                      & \textbf{$\Pi_{\sf nirank}$}       &       159                     \\
\multirow{2}{*}{Server}                   & \textbf{$\Pi_{\sf irank}$}        &      $\dfrac{99N}{J} + \big(91+71(J-1) \big) \cdot (\lceil \log ||R|| \rceil-1)$                                    \\
                                          & \textbf{$\Pi_{\sf nirank}$}       &   $\dfrac{99N}{J}+ \big(335N + \dfrac{11 N}{J}+ 71(J-1) (N+1)\big) \cdot (\lceil \log ||R|| \rceil-1)$                      \\ \bottomrule 
\end{tabular}
}
\label{tab:computationcomparison}
\end{table*}

\subsection{Communication Overhead}

We measured the upper bounds of the proof sizes of different Paillier-based zero-knowledge proofs shown in Table \ref{tab:zkpcommunicationcost} as they dominate the whole communication costs. We stress that these zero-knowledge proofs can be combined to reduce total communication costs. Table \ref{tab:outcommunicationcomparison} and \ref{tab:incommunicationcomparison} respectively demonstrate the outbound and inbound communication cost comparison of the two protocols, where inbound and outbound costs refer to the data flowing in and flowing out. For the interactive protocol $\Pi_{\sf irank}$, at first, each ${\mathcal{U}}_i$ should submit 6.25 KB data for an encrypted input ${\mathcal{E}}(x_i)$ and an zkpRG$\{x_i\}$. Then in each round, ${\mathcal{U}}_i$ needs to submit 15 KB  data, which is negligible compared to the general 4GLTE upload speed 512KB (4Mbps)\footnote{4G LTE Speeds of Verizon:https://www.verizon.com/articles/4g-lte-speeds-vs-your-home-network/.}. Thus, $\Pi_{\sf irank}$ requires each ${\mathcal{U}}_i$ to submit $6.25+15 \cdot (\lceil \log ||R|| \rceil-1)$ KB data in total without optimisations. Using optimisation-II to halve the round complexity, each ${\mathcal{U}}_i$ may need to submit $6.25 + \dfrac{15 \cdot 3}{2} \cdot (\lceil \log ||R|| \rceil-1)$ KB data. On the other hand, each server only needs to provide 1.5 KB data of the final aggregated result and a zkpPD in each round. For the inbound costs, each user receives 0.25 KB of the decrypted aggregated result from the servers. In contrast, the servers split all the users' inputs and their communicated data in each round with optimisation-IV.

For the online phase of the non-interactive protocol $\Pi_{\sf nirank}$, each ${\mathcal{U}}_i$ only needs to submit constant 6.25KB input data throughout the protocol. In each round, each server must transfer 5.75N KB data for all the users. Considering N=1000 users, each server merely transmits 5.6 MB data in each round, which is fairly negligible compared to their gigabit bandwidths. Moreover, for the inbound costs, each user only receives 0.25KB of the final aggregated result, whereas each server receives the users' inputs and the communicated data from the other $J-1$ servers. 

\subsection{Computational Overhead}

We implemented the Paillier-based zero-knowledge proofs and measured their running time in milliseconds. We also measured the running time of the Paillier encryption and decryption, which respectively consume 11ms and $91+(J-1) \cdot 71$ms on average, where decryption essentially includes the generation of a zkpPD (91ms) by each server and the verification of $J-1$ zkpPDs $\big(71 \cdot (J-1) {\rm ms}\big)$. We dismissed the running time of additive operations on ciphertexts as each additive operation consumes less than 1ms, which is computationally negligible in contrast to other operations. Table \ref{tab:zkpcomputationalcost} illustrates the computational costs of zero-knowledge proofs.

We estimated the computational costs of the two protocols in Table \ref{tab:computationcomparison}. For $\Pi_{\sf irank}$, at first, it takes 159 ms for each ${\mathcal{U}}_i$ to produce an encrypted input ${\mathcal{E}}(x_i)$ and a zkpRG$\{x_i\}$. Then in each round, ${\mathcal{U}}_i$ must spend 413 ms to generate all the necessary NIZKPoKs without optimisation. Besides, with optimisation-II, each ${\mathcal{U}}_i$ may need to spend $159 + \dfrac{413 \cdot 3}{2} \cdot (\lceil \log ||R|| \rceil-1)$ ms in total for computation. On the other hand, the servers split all the users' range proofs and each of them should spend $\dfrac{99N}{J}$ ms using a single processor. Then each server spends $91+(J-1)\cdot 71$ ms to generate a zkpPD and verify the other $J-1$ servers' zkpPDs using one processor. Concretely, for $||R||=2^8$, $N=1000$ and $J=8$, it takes each server about 17 seconds to process all the data.

For $\Pi_{\sf nirank}$, it only costs each ${\mathcal{U}}_i$ 159 ms to generate the input and a range proof in total. In each round, with a single processor, each server must spend $335N + \dfrac{11 \cdot N}{J}+ (J-1)\cdot 71\cdot (N+1)$ ms to generate $2N$ zkpMTPs, perform joint decryptions of $N$ users' encrypted inputs, $\dfrac{N}{J}$ re-encryptions and a joint decryption of the final aggregated result. More concretely, for $||R||=2^8$, $N=1000$ and $J=8$, each server needs about 1.6 hours to process all the data, which is substantially greater than the running time of $\Pi_{\sf irank}$. However, if each server uses multiple processors, our non-interactive protocol's running time would be considerably reduced. For instance, with 32 processors, it only costs each server about 3 minutes.

\section{Conclusion}

In this paper, we presented a practically efficient model for securely computing rank-based statistics over distributed datasets in the malicious setting. Based on the binary search technique, we proposed an interactive protocol and a non-interactive protocol to compute the $k^{th}$ element securely. Compared to the state-of-the-art solutions, our computing model provides stronger security and privacy while maintaining high efficiency and accuracy. It involves $O(N \log ||R||)$ time complexity, where $||R||$ is the range size of the dataset elements. Finally, we offered a complete and self-contained UC-secure instantiation of our computing model using the threshold Paillier cryptosystem and $\Sigma$-protocol zero-knowledge proofs of knowledge. Our performance evaluation corroborated the high accuracy and efficiency that our computing model can achieve.

\bibliographystyle{plain}
\bibliography{references}

\appendices
\section{Zero-knowledge Proof of Knowledge} \label{sec:append1}

We provide the Paillier-based $\Sigma$-protocol zero-knowledge proofs of knowledge. These proofs have completeness, soundness and special honest-verifier zero-knowledge properties. Note that the symbols and notations used in this section are independent of those in the main part. 

\subsection{Zero-knowledge Proof of Multiplication} 

Given ${\mathcal E}(x)$, ${\mathcal E}(y)=g^y \cdot \gamma^n$ and ${\mathcal E}(z)={\mathcal E}(x)^y \cdot \nu^n$, ${\mathcal{P}}$ can convince ${\mathcal{V}}$ of the knowledge that ${\mathcal E}(z)$ encrypts $z=x \cdot y ~(\text{mod}~n)$ without knowing $x$:
\begin{enumerate}
\item
${\mathcal{P}}$ randomly generates $m, \theta, \nu \in Z_n^*$ before sending ${\mathcal E}(m)=g^m \cdot \theta^n$ and ${\mathcal E}(xm)={\mathcal E}(x)^m \cdot \lambda^n$ to ${\mathcal{V}}$.

\item
${\mathcal{V}}$ sends a random challenge $e \xleftarrow{\$} Z_n^*$ to ${\mathcal{P}}$.

\item
${\mathcal{P}}$ replies with $p=m+e \cdot y~(\text{mod}~n)$, $w=\theta \cdot \gamma^{e}~(\text{mod}~n^2)$ and $u=\lambda \cdot {\mathcal{E}}(x)^t \cdot \nu^{e}~(\text{mod}~n^2)$.

\end{enumerate}
${\mathcal{V}}$ accepts if and only if $g^p \cdot w^n \overset{\sf ?}{=}  {\mathcal E}(m) \cdot {\mathcal E}(y)^{e}$ and ${\mathcal E}(x)^p \cdot u^n\overset{\sf ?}{=} {\mathcal E}(xm) \cdot {\mathcal E}(z)^{e}$.

Please refer to the detailed protocols and security proofs in Section 8.1.3 of \cite{MCTHE}.

\subsection{Zero-knowledge Proof of Membership}  

We adapted the state-of-the-art zero-knowledge polynomial evaluation protocol of Flashproofs (ASIACRYPT \textquotesingle 22) \cite{Flashproofs} over cyclic groups for Paillier-based membership proof. We only demonstrate the case where the witness $x$ is $-1$ or $1$ as below. Given ${\mathcal E}(x)=g^x \cdot \gamma^n$, ${\mathcal P}$ can convince ${\mathcal V}$ of $x \in \{-1, 1\}$:

\begin{enumerate}
\item ${\mathcal P}$ first randomly generates $(m, \lambda, \theta, \nu) \in {\mathbb Z}_n^*$. Then ${\mathcal P}$ computes and sends ${\mathcal E}(m)=g^m \cdot \lambda^n$, ${\mathcal{E}}(2mx)=g^{2mx} \cdot \theta^n$ and ${\mathcal{E}}(m^2)=g^{m^2} \cdot \nu^n$ to ${\mathcal V}$.

\item ${\mathcal V}$ sends a random challenge $e \xleftarrow{\$} {\mathbb Z}_n^*$ to ${\mathcal P}$. 

\item ${\mathcal P}$ replies with $p=m+ e\cdot x~(\text{mod}~n)$, $w=\lambda \cdot \gamma^e~(\text{mod}~n^2)$ and $u=\nu \cdot \theta^e~(\text{mod}~n^2)$ to ${\mathcal V}$.
\end{enumerate}
${\mathcal{V}}$ accepts if and only if $g^p \cdot w^n \overset{\sf ?}{=}  {\mathcal E}(m) \cdot {\mathcal E}(x)^e$ and $g^{(p^2-e^2)} \cdot u^n\overset{\sf ?}{=} {\mathcal E}(m^2) \cdot {\mathcal E}(2mx)^e$.

\subsection{Zero-knowledge Proof of Range} 

We adapted the state-of-the-art zero-knowledge range protocol (EUROCRYPT \textquotesingle 21) \cite{couteau} over cyclic groups for Paillier-based range proof. The protocol is based on the principle that any non-negative value $4x(B-x)$ for $x \in [0, B]$ can be decomposed to the sum of three squares. Please refer to the detailed protocols and security proofs in \cite{couteau}, where $B$ is the upper bound. Given ${\mathcal{E}}(x)=g^x \cdot r^n$, ${\mathcal P}$ can convince ${\mathcal V}$ of $x \in [0, B]$, where $r \xleftarrow{\$} {\mathbb{Z}_n}^*$:
\begin{enumerate}
\item ${\mathcal{P}}$ computes $4x(B-x)+1=\sum_{i=1}^3 x_i^3$ and sets $x_0=B-x,~r_0=r^{-1}$. Then ${\mathcal{P}}$ randomly generates $(r_i \xleftarrow{\$} {\mathbb{Z}_n}^*, m_i \xleftarrow{\$} (0, BCL], s_i \xleftarrow{\$} {\mathbb{Z}_n}^*)_{i=1}^3,~\rho \xleftarrow{\$} {\mathbb{Z}_n}^*$ and computes:
\begin{itemize}
\item${\mathcal{E}}(x_0)=g^B \cdot {\mathcal{E}}(x)^{-1}$

\item$\big({\mathcal E}(x_i)=g^{x_i} \cdot r_i^n\big)_{i=1}^3$

\item$\big({\mathcal E}(m_i)=g^{m_i} \cdot s_i^n\big)_{i=0}^3$

\item$d={\mathcal{E}}(x)^{4m_0} \cdot \prod_{i=1}^3 {\mathcal{E}}(x_i)^{-m_i} \cdot \rho^n$.

\item$\Delta={\mathcal{H}}\big({\mathcal{E}}(m_0), {\mathcal{E}}(m_1), {\mathcal{E}}(m_2), {\mathcal{E}}(m_3), d\big)$

\end{itemize}
where $L$ is the growth factor of masked intervals due to additive noise and $C$ is the upper bound of the challenge.

${\mathcal{P}}$ sends $\big({\mathcal E}(x_i)\big)_{i=0}^3$, $d$ and $\Delta$.

\item
${\mathcal{V}}$ sends a random challenge $e \xleftarrow{\$} [0,C]$ to ${\mathcal{P}}$.

\item
${\mathcal{P}}$ replies with $\big(p_i=m_i+e\cdot x_i~(\text{mod}~n)\big)_{i=0}^3$, $\big(w_i=s_i \cdot r_i^e~(\text{mod}~n^2)\big)_{i=0}^3$ and $\tau=\rho \cdot \prod_{i=1}^3 r_i^{e x_i} \cdot r^{-4e x_0}~(\text{mod}~n^2)$.

\end{enumerate}
${\mathcal{V}}$ accepts if and only if:
\begin{enumerate}
\item $(f_i=g^{p_i} \cdot w_i^n \cdot {\mathcal{E}}(x_i)^{-e})_{i=0}^3$

\item $f=\tau^n \cdot g^e \cdot {\mathcal{E}}(x)^{4p_0} \cdot \prod_{i=1}^3 {\mathcal{E}}(x_i)^{-p_i}$

\item $\Delta \overset{\sf ?}{=} {\mathcal{H}}\big(f_0, f_1, f_2, f_3, f\big)$ 

\item $(p_i \in [0, BC(L+1)])_{i=0}^3$

\end{enumerate}

\subsection{Zero-knowledge Proof of Non-Zero}

Given ${\mathcal{E}}(x)$, ${\mathcal{P}}$ can convince ${\mathcal{V}}$ of the knowledge that $x \neq 0$. ${\mathcal{P}}$ encrypts $y=x^{-1}$ as ${\mathcal{E}}(y)$ and proves $x \cdot y = 1~(\text{mod}~n)$:
\begin{enumerate}
\item
${\mathcal{P}}$ randomly generates $y=x^{-1}, r_y, m, r_m, v \in Z_n^*$ before sending ${\mathcal E}(y)=g^y \cdot r_y^n$, ${\mathcal E}(m)=g^m \cdot r_m^n$ and ${\mathcal E}(xm)={\mathcal E}(x)^m \cdot v^n$ to ${\mathcal{V}}$.

\item
${\mathcal{V}}$ sends a random challenge $e \xleftarrow{\$} Z_n^*$ to ${\mathcal{P}}$.

\item
${\mathcal{P}}$ replies with $p=m+e \cdot y~(\text{mod}~n)$, $w=r_m \cdot r_y^e~(\text{mod}~n^2)$ and $u=r_x^{tn-ey} \cdot v~(\text{mod}~n^2)$, where $t$ is defined by $m+c\cdot y=p+tn$.

\end{enumerate}
${\mathcal{V}}$ accepts if and only if $g^p \cdot w^n \overset{\sf ?}{=}  {\mathcal E}(m) \cdot {\mathcal E}(y)^e$ and ${\mathcal E}(x)^p \cdot u^n\overset{\sf ?}{=} {\mathcal E}(xm) \cdot g^e$.

Please refer to the detailed protocols and security proofs in Section 8.1.3 of \cite{MCTHE}.

\subsection{Zero-knowledge Proof of Partial Decryption} 

Given a public verification key pair ($v, v_i$) for the $i$-th private key ${\sf sk_i}$, ${\mathcal P}$ can convince ${\mathcal V}$ that ${\mathcal E}(x)_i={\mathcal E}(x)^{2 \cdot \Delta \cdot {\sf sk_i}}$ is produced with her secret key ${\sf sk_i}$, where $v_i=v^{\Delta \cdot {\sf sk_i}}~(\text{mod}~n^2)$ and $\Delta=N!$ for $i \in \{1, ..., N\}$:

\begin{enumerate}
\item ${\mathcal P}$ randomly generates $r \in {\mathbb Z}_n^*$ before sending ${\mathcal E}(x)^{4r}$ and $v^r~(\text{mod}~n^2)$ to ${\mathcal V}$.

\item ${\mathcal V}$ sends a random challenge $e \in {\mathbb Z}_n^*$ to ${\mathcal P}$.

\item ${\mathcal P}$ replies with $p=r+e \cdot \Delta \cdot {\sf sk_j}~(\text{mod}~n)$.
\end{enumerate}
${\mathcal V}$ accepts if ${\mathcal E}(x)^{4p}\overset{\sf ?}{=} {\mathcal E}(x)^{4r} \cdot {\mathcal E}(x)_i^{2 \cdot e}$ and $v^p \overset{\sf ?}{=} v^r \cdot v_i^e$.

Please refer to the detailed protocols and security proofs in Section 5.1 of the Damg{\aa}rd-Jurik cryptosystem \cite{pailliergeneralization}.

\section{Distributed Decryption Protocol $\Pi_{\sf ddec}$} \label{sec:append2}

This protocol allows a group of distributed parties to perform the decryption of a Paillier ciphertext jointly. Given a ciphertext, the parties broadcast the partial decryption shares and zero-knowledge proofs of partial decryption (zkpPD). Then the parties check the proofs and collect the partial shares to obtain the plaintext. It is trivial to build a simulator to prove UC security. With the power of controlling the random oracle, the simulator can simulate the zkpPDs and decrypt the ciphertext to any desired value.

\section{Secret-Sharing Protocol $\Pi_{\sf sr}$} \label{sec:append3}

We adapted the UC-secure reshare protocol proposed in the commitment-enhanced SPDZ protocol \cite{cespdz}. A group of $N$ parties secretly share a Paillier ciphertext ${\mathcal{E}}(x)$, such that the $i$-th party ${\mathcal{P}}_i$ holds a share ${\mathcal{E}}(x_i)$, where $x=\sum_{i=1}^N x_i$:

\begin{enumerate}
\item ${\mathcal P}_i$ generates random values $v_i, r_i \in {\mathbb{Z}_n}^*$ to create a ciphertext ${\mathcal{E}}({v_i})=g^{v_i} r_i^n$.

\item ${\mathcal P}_i$ computes ${\mathcal{E}}(x+v)={\mathcal{E}}(x) \cdot \prod_{i=1}^N {\mathcal{E}}(v_i)$, where $v=\sum_{i=1}^N v_i$.

\item The parties obtains $x+v$ by jointly decrypting ${\mathcal{E}}(x+v)$ with $\Pi_{\sf ddec}$. Then they jointly generate a random value $r_{x+v} \in {\mathbb{Z}_n}$. 

\item ${\mathcal{P}}_1$ locally obtains $x_1=x+v-v_1$ and $r_{x_1}=r_{x+v} \cdot r_1^{-1}$ whereas ${\mathcal{P}}_i, i \in \{2,..., N\}$ can get $x_i=-v_i$ and $r_{x_i}=r_i^{-1}$.

\item ${\mathcal P}_i$ locally obtains a share ${\mathcal{E}}(x_i)$ by computing ${\mathcal{E}}(x_1)={\mathcal{E}}(x+v) \cdot {\mathcal{E}}(v_i)^{-1}$ and ${\mathcal{E}}(x_i)={\mathcal{E}}(v_i)^{-1}, i \in \{2, ..., N\}$. 

\end{enumerate}

\section{Pre-processing Protocol $\Pi_{\sf prep}$} \label{sec:append4}

This protocol allows a group of parties $({\mathcal{P}}_i)_{i=1}^N$ to jointly generate a sufficient number of encrypted random values ${\mathcal{E}}(r)$ and ${\mathcal{E}}\big(\phi(r)\big)$. The parties must perform the computations alternately using the additively homomorphic Paillier cryptosystem with an RSA modulus $n$. That is to say, the party ${\mathcal{P}}_{i>1}$ receives the results from the previous party ${\mathcal{P}}_{i-1}$, performs the computations and sends the results to the next party ${\mathcal{P}}_{i+1}$. Let $\Gamma=\big[-(\frac{n-1}{16(\overline{B}-\underline{B})})^{\frac{1}{N}}, 0\big) \cup \big(0,(\frac{n-1}{16(\overline{B}-\underline{B})})^{\frac{1}{N}}\big]$ be the randomness space. The protocol starts with the party ${\mathcal{P}}_1$:

${\mathcal{P}}_1$ encrypts a random integer $r_1 \in \Gamma$ and its sign $\phi(r_1)$ as ${\mathcal{E}}(r_1)$ and ${\mathcal{E}}\big(\phi(r_1)\big)$. Then ${\mathcal{P}}_1$ sends the ciphertexts to ${\mathcal{P}}_2$ along with a tuple of zero-knowledge proofs that satisfy the relations defined for the language ${\mathcal{L}}_2$ in Section \ref{sec:preprocessing}:
\begin{itemize}
\item zkpRG$\{{\mathcal{E}}(r_1), \Gamma\}$, zkpNZ$\{{\mathcal{E}}(r_1), r_1 \neq 0\}$

\item zkpMBS$\big\{{\mathcal{E}}\big(\phi(r_1)\big), \{-1, 1\}\big\}$

\item zkpMTP$\{{\mathcal{E}}(r_1), {\mathcal{E}}\big(\phi(r_1)\big), {\mathcal{E}}\big(\phi(r_1) \cdot r_1\big)\}$

\item zkpRG$\{{\mathcal{E}}\big(\phi(r_1) \cdot r_1\big), [0, (\frac{n-1}{16(\overline{B}-\underline{B})})^{\frac{1}{N}}]\}$

\end{itemize}
Apart from ${\mathcal{P}}_1$'s results, ${\mathcal{P}}_2$ needs to produce two zkpMTPs, namely, a zkpMTP$\{{\mathcal{E}}(r_1), {\mathcal{E}}(r_2), {\mathcal{E}}(r_1 r_2)\}$ and a zkpMTP$\{{\mathcal{E}}\big(\phi(r_1)\big), {\mathcal{E}}\big(\phi(r_2)\big), {\mathcal{E}}\big(\phi(r_1) \phi(r_2)\big)\}$. Next, ${\mathcal{P}}_2$ sends them to ${\mathcal{P}}_3$, who would perform the same computations as ${\mathcal{P}}_2$. Finally, all the parties obtain the two encrypted values ${\mathcal{E}}(\prod_{i=1}^N r_i)$ and ${\mathcal{E}}\big(\prod_{i=1}^N \phi(r_i)\big)$. The parties jointly run the protocol $\Pi_{\sf sr}$ to secretly share these values such that each party holds a partial share. Note that the zero-knowledge proofs can be combined into one proof by using only one random challenge.

We briefly describe a non-rewindable UC simulation. The simulator ${\mathcal{S}}$ firstly calls ${\mathcal{F}}_{\sf keygen}$ to obtain the secret decryption keys. ${\mathcal{S}}$ is able to obtain all the plaintext inputs provided by the adversary ${\mathcal{A}}$. For the honest parties, ${\mathcal{S}}$ provides the simulated ciphertexts and the simulated NIZKPoKs, which are computationally indistinguishable to ${\mathcal{Z}}$. ${\mathcal{S}}$ aborts the protocol if any NIZKPoKs provided by ${\mathcal{Z}}$ are invalid.

\end{document}